%% file: circular_sgrastar_revision.tex
\documentclass[apj]{emulateapj}
\pdfoutput=1
\usepackage{graphicx}
 \usepackage{subfigure}
\usepackage{amssymb,amsmath}

\shorttitle{CIRCULAR POLARIZATION OF SGR A*}
\shortauthors{Mu\~noz et al.}
\slugcomment{Accepted for Publication in the Astrophysical Journal}
\newcommand{\sgr}{Sgr~A*}
\newcommand{\pasa}{PASA}

\begin{document}
\title{The Circular Polarization of Sagittarius A*\\
 at Submillimeter Wavelengths}

\author{D. J. Mu\~noz\altaffilmark{1}, D. P. Marrone\altaffilmark{2,3,4}, J. M. Moran\altaffilmark{1}, and
R. Rao\altaffilmark{5} }
\affil{{$^1$}Harvard-Smithsonian Center for Astrophysics, 60 Garden Street, Cambridge, MA 02138\\
{$^2$}Kavli Institute for Cosmological Physics, University of Chicago, 5640
South Ellis Avenue, Chicago, IL 60637\\
{$^3$}Steward Observatory, University of Arizona, 933 North Cherry Avenue, Tucson, AZ 85721\\
{$^4$}Hubble Fellow\\
{$^5$}Submillimeter Array, Academia Sinica Institute of Astronomy and Astrophysics,\\
 645 N. Aohoku Place, Hilo, HI 96720}

\email{dmunoz@cfa.harvard.edu\\ \today}

\begin{abstract}

We report the first detections of circularly polarized emission at submillimeter wavelengths from the compact radio source and supermassive black hole candidate \sgr\ 
at a level of $1.2\pm0.3\%$ at 1.3 mm wavelength (230~GHz) and $1.6\pm0.3\%$ at 860~$\mu$m (345~GHz) with the same handedness, left circular polarization (LCP), as observed at all lower frequencies (1.4--15~GHz). 
The observations, taken with the Submillimeter Array in multiple epochs, also show simultaneous linear polarization (LP) at both wavelengths of about 6\%. These properties differ sharply from those at wavelengths longer than 1~cm (frequencies below 30~GHz), where weak circular polarization (CP) ($\sim0.5\%$) dominates over LP, which is not detected at similar fractional limits. We describe an extensive set of tests to ensure the accuracy of our measurements. We find no circular polarization (CP) in any other source, including the bright quasar 1924-292, which traces the same path on the sky as \sgr\ and therefore should be subject to identical systematic errors originating in the instrument frame. Since a relativistic synchrotron plasma is expected to produce little CP, the observed CP is probably generated close to the event horizon by the Faraday conversion process. We use a simple approximation to show that the phase shift associated with Faraday conversion can be nearly independent of frequency, a sufficient condition to make the handedness of CP independent of frequency. Because the size of the $\tau=1$ surface changes by more than an order of magnitude between 1.4 and 345~GHz, the magnetic field must be coherent over such scales to consistently produce LCP. To improve our understanding of the environment of SgrA* critical future measurements include determining whether the Faraday rotation deviates from a $\lambda^2$ dependence in wavelength and whether  the circular and linear components of the flux density are correlated.

\end{abstract}

\keywords{black hole physics --- Galaxy: center --- plasmas --- polarization --- submillimeter ---
techniques: interferometric}

\section{Introduction} 

The Galactic Center source Sagittarius A* (Sgr~A*), the nearest supermassive black hole (SMBH),
is extremely underluminous for its mass \citep[$\sim4\times10^6$ M$_\odot$,][]{ghe08,gil09}, radiating at only $10^{-9}\,L_\mathrm{Edd}$. Theoretical models have focused their efforts on explaining this faintness by invoking diverse physical mechanisms that result in radiatively
inefficient accretion and/or outflow processes \citep{fal93, nar94,bla99,qua00a,fal00,yua02}. 
These models adequately reproduce the quiescent spectrum of \sgr\ \citep{nar98,mar01,mel01b,yua03}, although the spectrum alone does not discriminate between them. 

In subsequent years, new observations have provided new types of constraints on the models. Very long baseline interferometry (VLBI) has measured the wavelength-dependent size of \sgr, detecting structure on event-horizon scales in the highest frequency/resolution experiments \citep{she05,bow06,doe08,fis11}. The size-wavelength relation, and even the observed interferometric visibility of the emission, can be matched by several models
\citep{mar07,mos09,hua09b,bro10}. Multiwavelength variability \citep{marr08b,yus08,kunn10}, in combination with other observables, which may also reveal information about source structure, can also be replicated in multiple schemes
\citep{dex09,mai09}.

\input{table_pol_b}      

The polarization spectrum and its variability present additional, rich information about source structure. Strong theoretical constraints can be derived by considering the effects of polarized radiative transfer (PRT) in the models.
Perhaps the simplest and most common consequence of PRT through a plasma is Faraday rotation of linear polarization (LP).  In recent years, multifrequency millimeter and submillimeter polarimetry has allowed measurements of Faraday rotation in \sgr \citep{marr06a,marr07}. These data have provided bounds on the accretion rate in the inner regions of the flow \citep{mac06,marr07}.

A synchrotron-emitting plasma can be expected to produce significant LP fractions at frequencies near the spectral peak. 
At the same time, these relativistic plasmas are expected to produce very small amounts of circular polarization (CP) by intrinsic emission of Stokes $V$\citep{lan75} or through various radiative transfer effects \citep[e.g.,][]{jon77a,mel97,rus02,shch08}. 
As the polarization arises very near to the black hole, a full picture of the emergent polarization state can diagnose both the inner regions and the intervening propagation medium.

\begin{figure*}[bt]  
\epsscale{1.0}
\plotone{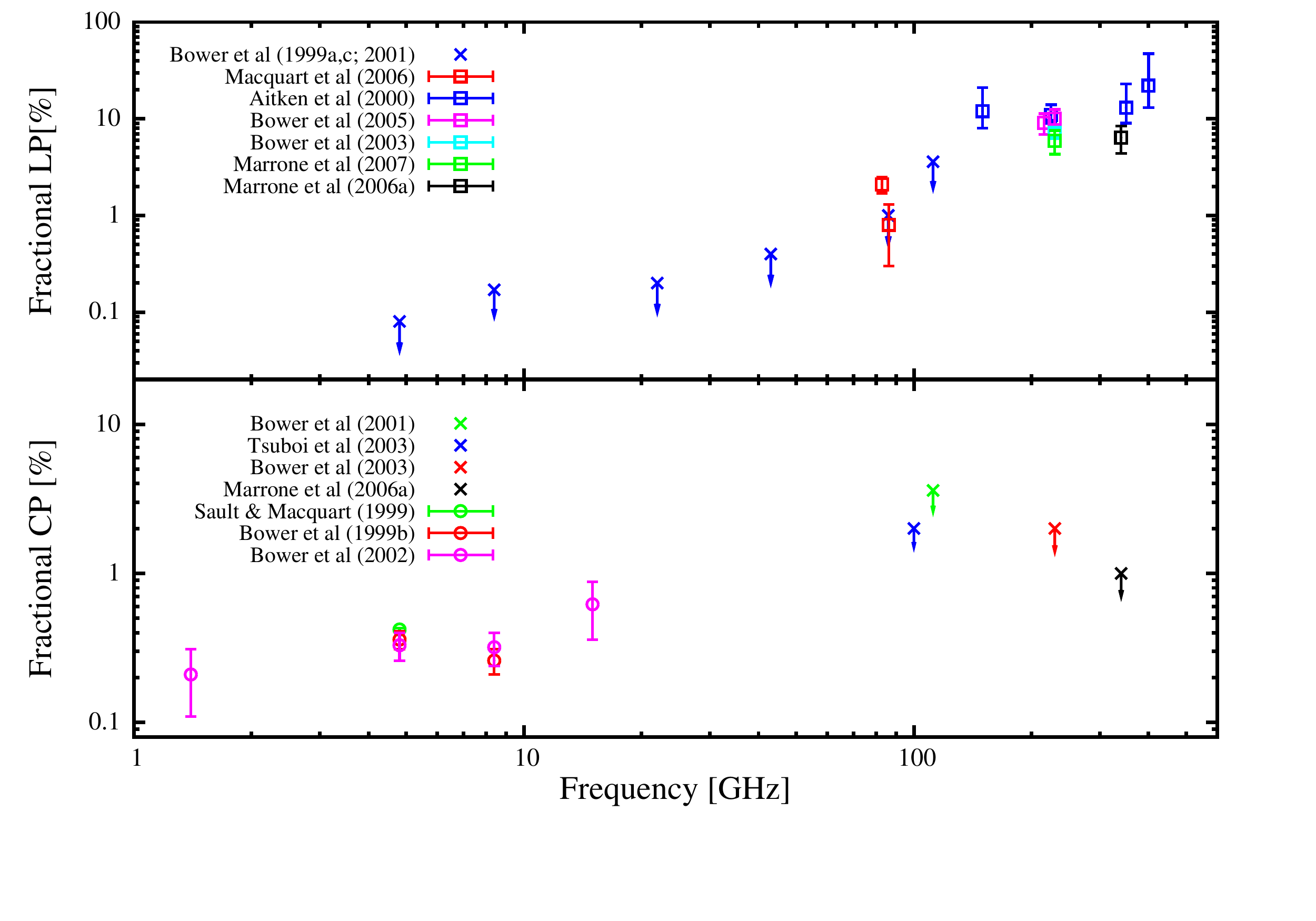}
    \caption{\label{fig:pol_data} 
Published measurements of the fractional linear and circular polarization toward Sgr~A* as a function of frequency.}
\end{figure*}

A compilation of published polarization measurements is given in Table~\ref{tab:table_pol} and shown in Figure~\ref{fig:pol_data}. No LP has been observed at
8 GHz or below at limits of 0.2\% or less \citep{bow99a}.
The LP of \sgr\ was first detected by \citet{ait00} above 100 GHz. These measurements were obtained with large beams ($7''$--$22''$) and required background subtraction, leaving some uncertainty about residual polarization contamination from the surrounding dust emission. 
Subsequent interferometer observations at millimeter and submillimeter wavelengths have shown polarization at the level of 1--10\%, which varies in position angle \citep{bow05} and fraction \citep{marr06a} 
with timescales comparable to those of previously observed total intensity variations \citep{marr06b}. The variability may be intrinsic to
the source or due to propagation effects, but the short timescales involved suggest that processes very close to the SMBH are responsible. The LP was measured simultaneously at multiple frequencies for the first time by \citet{marr07}. The inferred
rotation measure (RM)  indicates that \sgr\ is extremely underfed, with an accretion rate of $\sim10^{-8}\,M_\odot$~yr$^{-1}$.
 
CP from \sgr\ was first detected by \citet{bow99b}. Later, \citet{bow02} reported spectral measurements between 1.4 and 15 GHz as well as time variability of CP (see lower panel in Figure~\ref{fig:pol_data}). They concluded that
 the time-averaged CP spectrum is approximately $\nu^{0.5\pm 0.2}$, with persistent variability that increases with frequency. They also noted that the sense of CP for all available measurements (about 100 measurements from 1981 to 1999, made with both the VLA and ATCA at 1.4, 4.8, and 15 GHz)  was exclusively LCP, indicative of a long-term stability in magnetic field configuration.    At these frequencies, LP is not detected down to instrumental limits of 0.1\% while CP is persistently detected at levels of a fraction of a percent with a degree of variability of the same order of magnitude. This is substantially different from what is seen in high-luminosity active galactic nuclei (AGN), where LP always dominates CP. Above roughly 100~GHz, LP becomes dominant in \sgr, with CP undetectable at $\sim$1\% sensitivity.
In four epochs of 230~GHz observations, \citet{bow03} found a 2-$\sigma$ CP signal (3\%) on one day, with an average measurement of $1\%\pm1\%$. No CP was observed at a level of 0.5\% in the 100~GHz data reported by \citet{tsu03}. \citet{marr06a} measured $V$ at 345~GHz and obtained $-0.5\pm0.3\%$. However, because of systematic calibration uncertainties, they reported this result as an upper limit of $1.5\%$.
  
CP is also observed in a variety of radio sources, including pulsars and AGN. Examples are 3C273 and 3C279 \citep{hom99}. Some models seeking to explain the millimeter and submillimeter LP have also predicted CP at these high frequencies due to the conversion of LP to CP in a turbulent jet \citep[e.g.,][]{beck02}.  In these models, in addition to the stochasticity of the magnetic field---which appears to play a crucial role in building up CP by propagation effects---the helical geometry of jets might be important in high levels of LP, above 100 GHz. 

Coupled mechanisms to produce both LP and CP in relativistic outflows have been studied in detail by many authors
\citep[e.g.,][]{rus02,beck02,beck03,hua08,hom09,shch10b}. If CP is produced predominantly by propagation effects, multifrequency measurements of RM and CP can provide important clues about the magnetic field structure of the plasma surrounding the SMBH. Similarly, simultaneous measurements of LP and CP variability can determine whether the intraday variability in Sgr~A* is due to intrinsic variations in the central engine or to variations in the outer layers of the accretion flow. In this way, it should be possible to conclusively infer the presence or absence of a ``Faraday screen" in front of Sgr~A* \citep{mac06,marr07}.

\input{tab_tracks2}      

In Section~\ref{sec:observations} we describe several epochs of polarimetric observations of \sgr, Section~\ref{sec:results} reports the observed polarization, and Section~\ref{sec:errors} explores the tests of systematic errors in the observations that might give rise to false CP. Section~\ref{sec:cp_sources} reviews PRT and the astrophysical mechanisms for generating CP, and their applicability to \sgr.

\section{Observations}\label{sec:observations}

\input{tab_sgra_pol}    

The data presented in this work consist of three polarimetry tracks (Table~\ref{tab:tracks2}) taken with the Submillimeter Array (SMA).\footnote{The Submillimeter 
Array is a joint project between the Smithsonian Astrophysical Observatory 
and the Academia Sinica Institute of Astronomy and Astrophysics and is 
funded by the Smithsonian Institution and the Academia Sinica.} The general characteristics of the SMA are described by \citet{blu04} and  \citet{ho04}, and its polarimeter is described by  \citet{marronethesis} and \citet{marr08a}. 
The most significant detection of CP at 230~GHz was made from the observations of March 31, 2007. Archival data from two other observations---one at 230~GHz and another at 345~GHz---were analyzed to confirm and extend the initial result. A typical Sgr~A* polarimetry track consisted of \hbox{$\sim12$} hours of observations, of which \hbox{$\sim6$} corresponded to continuous monitoring of Sgr~A*. The quasar 1733-130 was used as the gain calibrator of Sgr~A*, while quasars 3C273 and 3C279 were used to calibrate for instrumental polarization (leakage). In addition, quasars such as 1337-129 and 3C286 were commonly included in the tracks, as well as a solar system object  (e.g., Titan) for flux density calibration.

\begin{figure}[hb!]  
\epsscale{1.0}
\plotone{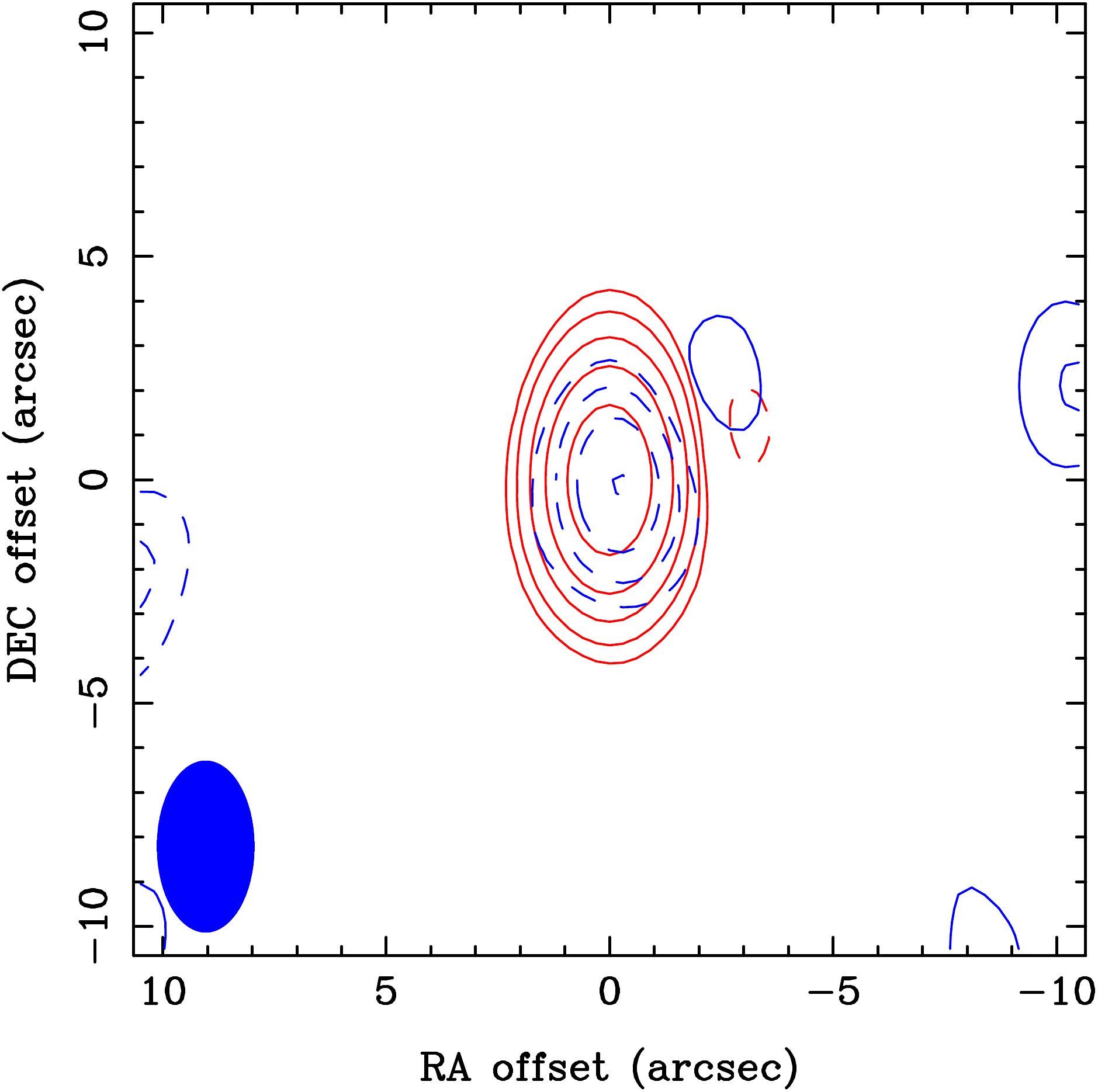}
\caption{Total and CP emission of Sgr~A* (March 31, 2007): red contours correspond to Stokes $I$ flux density while blue contours correspond to Stokes $V$ flux density. Solid and dashed contours indicate positive and negative flux density, respectively.  Stokes
$I$ contour spacings are $-6,6,12,25,50,100\,\times\sigma_I$, where $\sigma_I=19.8$~mJy/beam is the rms noise in the $I$ map. The contour levels for the Stokes $V$ map are $-11,-7,-4,-2,2,4\times\sigma_V$, where $\sigma_V=2.9$~mJy/beam is the rms noise in the $V$ map.
\label{fig:total_map}}
\end{figure}

To test the accuracy of our CP measurement, we made a special observation of the quasar 1924-292 at 230~GHz on May 30, 2008, along with another short-track measurement of \sgr. 1924-292 has nearly the same declination as Sgr~A*, so both have the same AZ--EL track and, hence, position angle dependence with time. This track also contained a one-hour segment on Sgr~A*. The gain calibrator for 1924-292 was 1911-201, and the polarization calibrators were 3C273 and 3C279. Every effort was made to keep the observational circumstances identical to those of previous tracks (e.g., same polarization sampling cycle).

The SMA polarimeter allows precise gain calibration for the $RR$ and $LL$ visibilities. A single quarter-wave plate is located in the beam of each telescope and rotates to produce either right CP or LCP. The polarization of each antenna is modulated according to orthogonal Walsh functions of period 16 to efficiently sample all four polarization combinations on each baseline. Quasi-simultaneous polarization data are generated in post-processing by averaging over the switching cycle. The use of the same waveplate and feed for both $R$ and $L$ polarization states eliminates some of the uncertainties inherent in the use of dual-feed receivers, particularly differential phase variations between the polarization states. Measurement of LP (Stokes $Q$ and $U$) relies on precise determination of the leakage of each polarization state by the crosshanded polarization, which was performed using linearly polarized
bright point sources (quasars 3C279 or 3C273) observed over a large range of parallactic angle. To first order, these leakages do not affect
the measurement of Stokes $V$ \citep[e.g.,][and Section~\ref{sec:gains}]{marronethesis,TMS}, leaving the relative calibration of the $RR$ and $LL$ visibilities as the primary calibration challenge for measurement of Stokes $V$.


\section{Results}\label{sec:results}

The polarization measurements (Stokes $I$, $Q$, $U$, and $V$) of Sgr~A* for four different epochs included in our analysis are shown in Table~\ref{tab:sgra_pol}. The flux densities for each Stokes parameter $I$, $Q$, $U$, and $V$ were obtained by fitting a point-source model to the visibilities, fixing the position to the phase center of the observations, the location of \sgr. The polarization Stokes parameters are listed as a fraction of the total flux density. For four different epochs---including different frequencies, different $uv$-space sampling, and a time span of three years---Sgr~A* is shown consistently to be 
circularly polarized at the $\gtrsim1\%$ level. The sense of CP (negative $V$) persists throughout.

 \begin{figure}[h]  
\epsscale{1.2}
\plotone{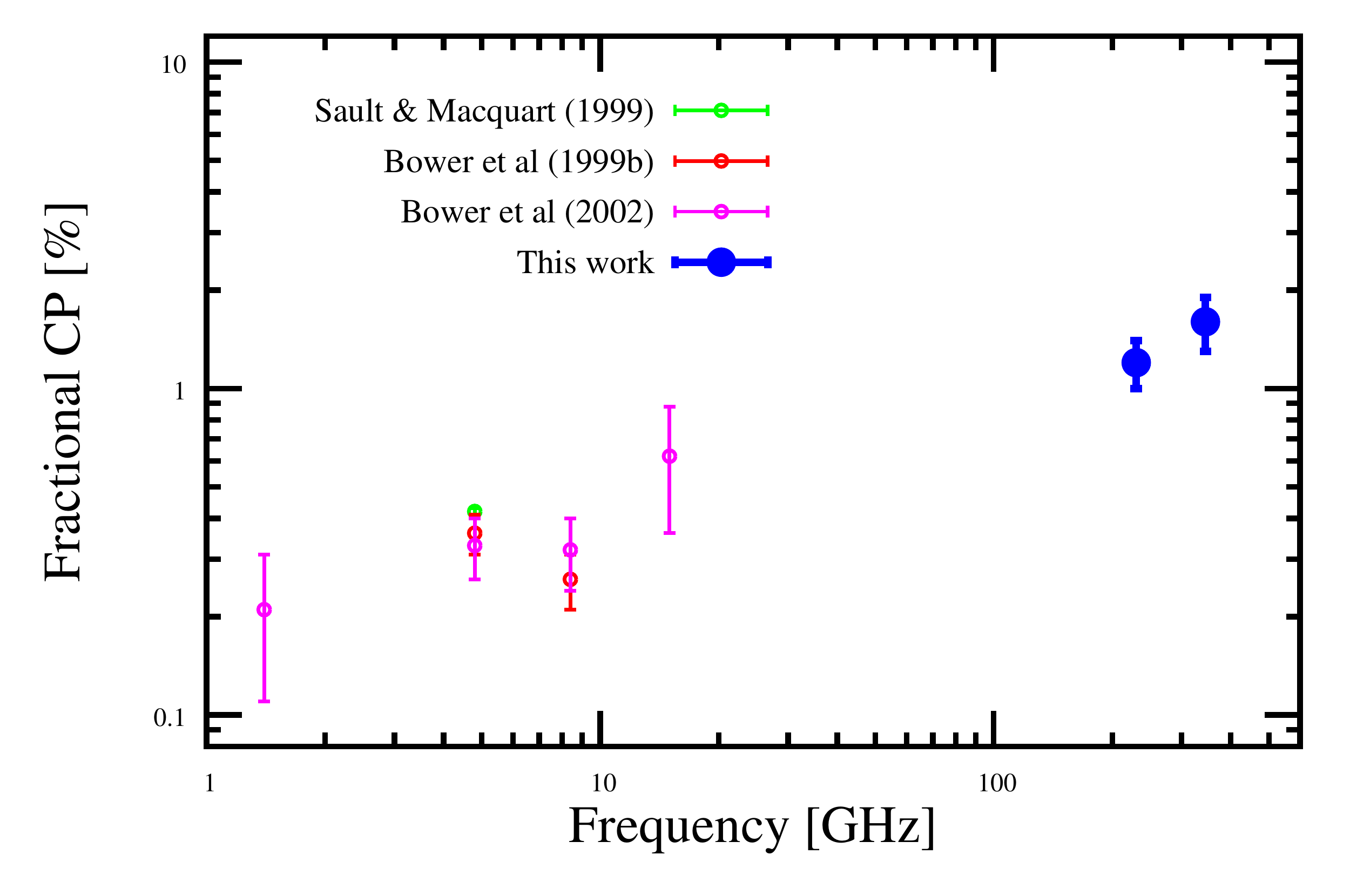}
    \caption{Fractional CP in Sgr~A* (LCP in all cases) from radio to submillimeter frequencies (lower panel in Figure~\ref{fig:pol_data}), including the new SMA data (see Table~\ref{tab:sgra_pol}). The percentage of CP increases with frequency  as $\sim\nu^{0.35\pm0.03}$.}
    \label{fig:pol_data_d}
\end{figure}

\begin{figure*}[ht!]  
\epsscale{0.35}
\begin{center}
\begin{tabular}[c]{cc}
  \begin{tabular}[c]{c}
	\plotone{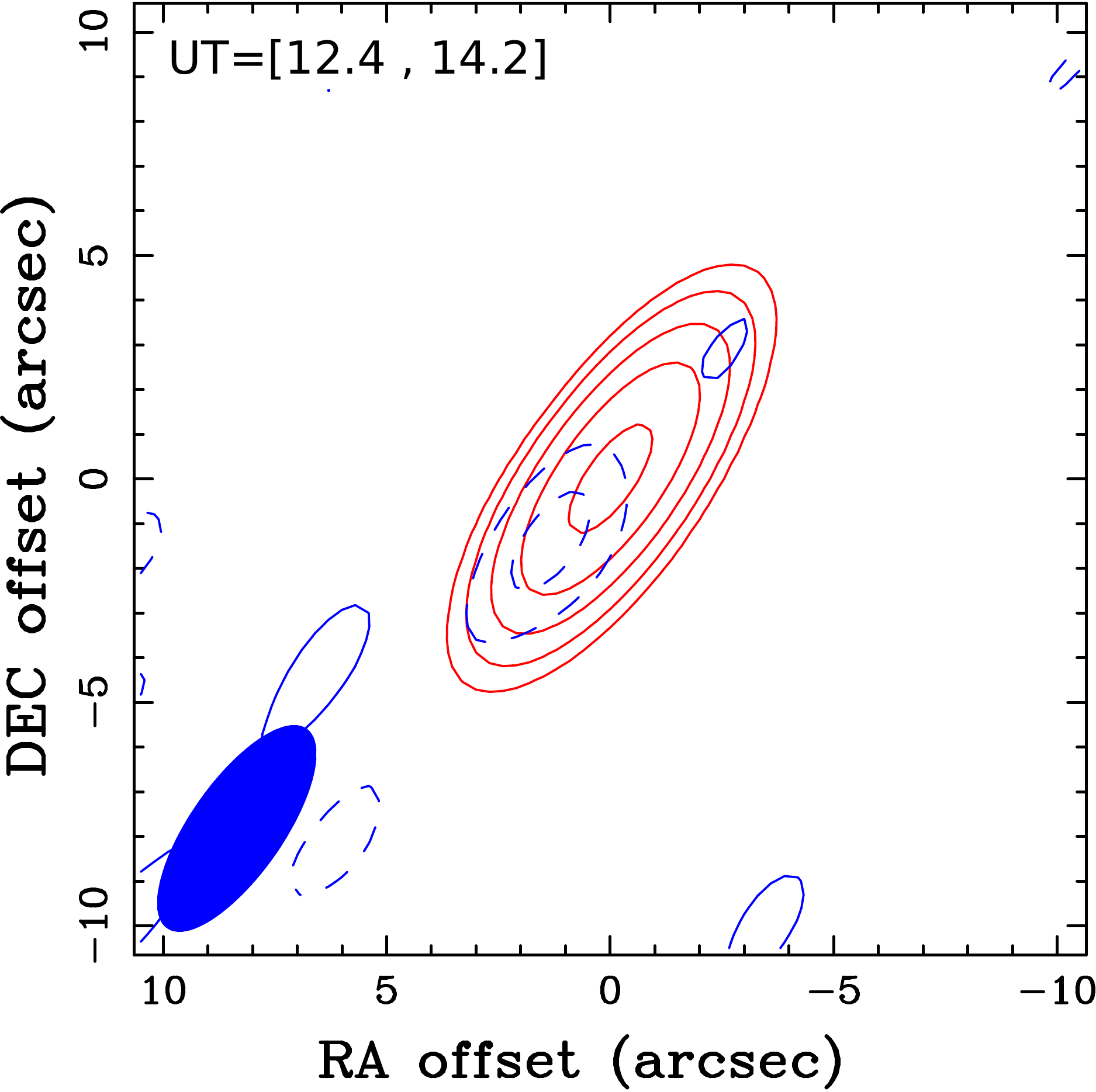}
  \end{tabular} &
  \begin{tabular}[c]{c}
	\plotone{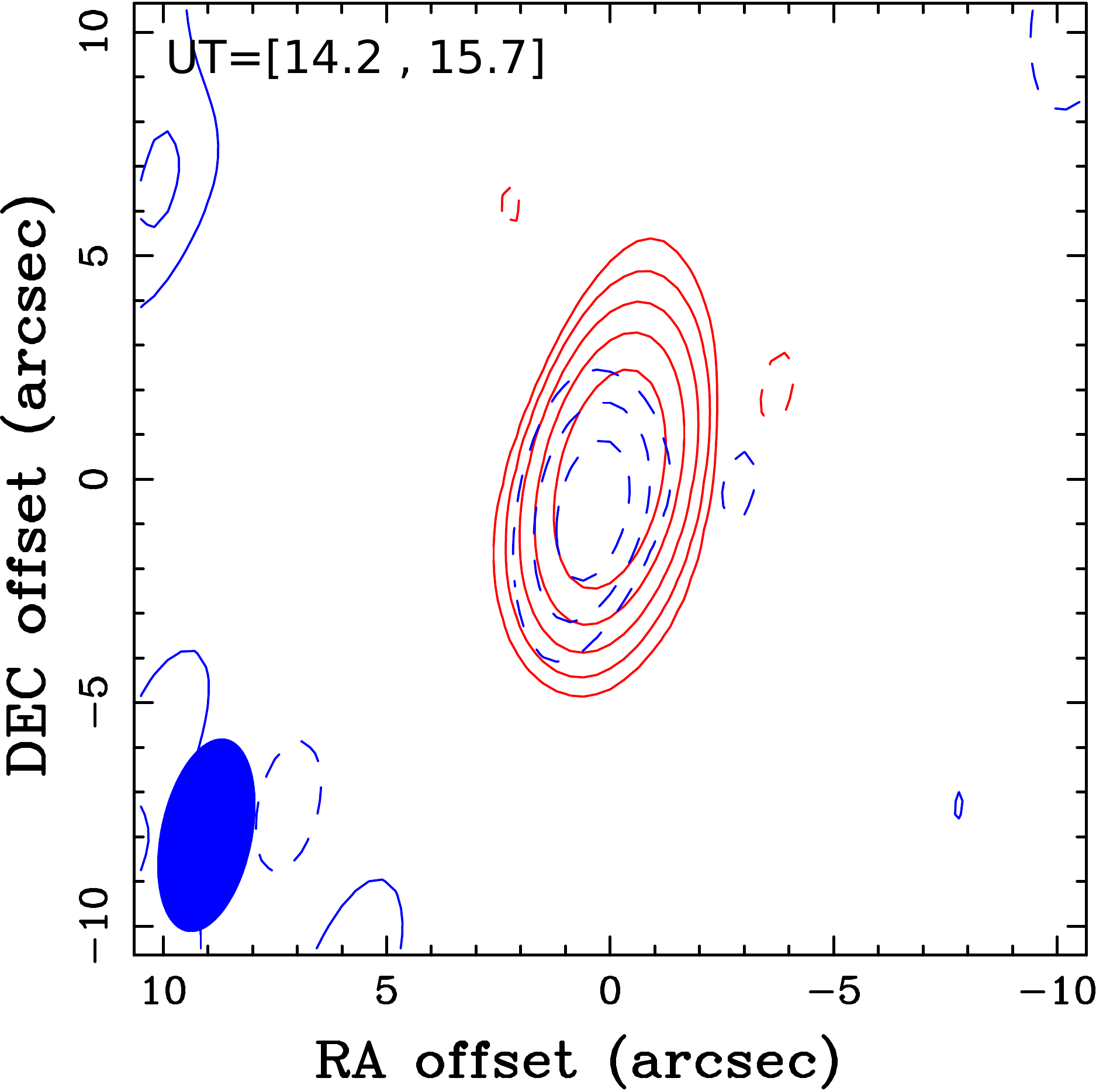}
  \end{tabular} \\
  \begin{tabular}[c]{c}
	\plotone{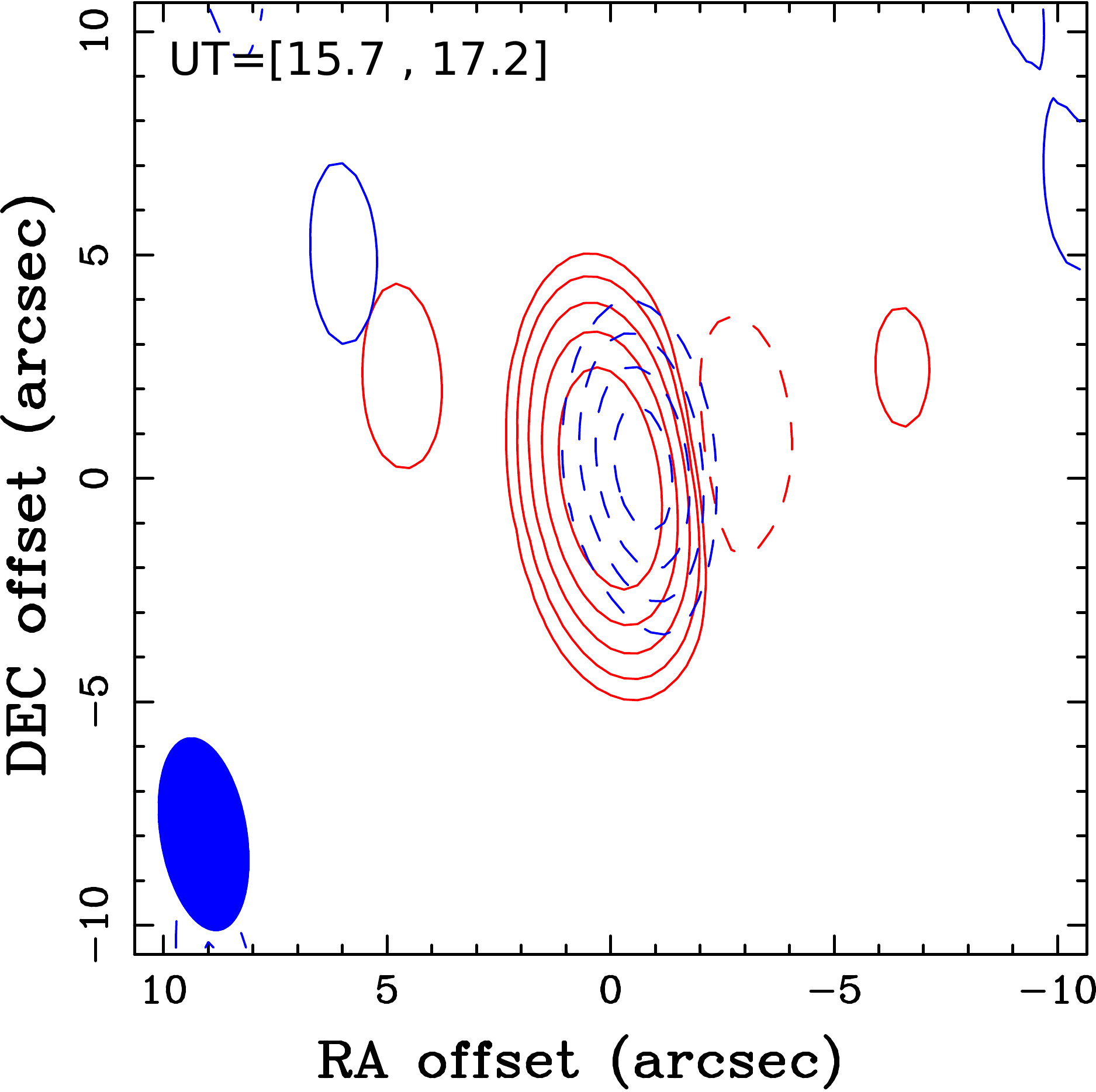}
  \end{tabular} &
  \begin{tabular}[c]{c}
	\plotone{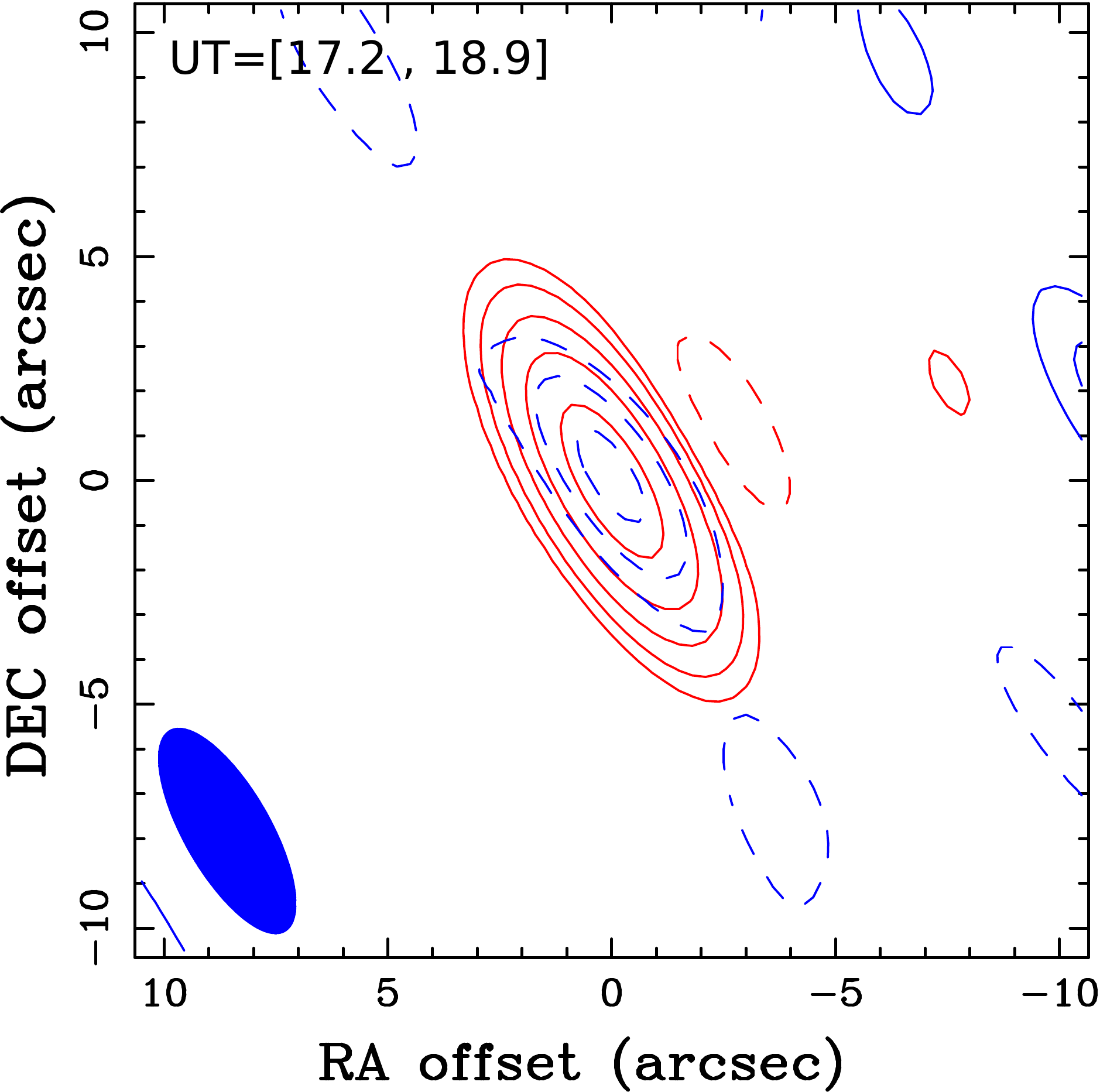}
  \end{tabular} \\
\end{tabular}
\end{center}
\caption{Contour images of Sgr~A* made from four segments of the track on March 31, 2007. Red: Stokes $I$ in Jy/beam. Blue: Stokes $V$ in Jy/beam. The color coding and contour levels are the same as in Figure~\ref{fig:total_map}. Each of the four panels corresponds to one of the four time intervals the track was divided into: UT=13.4, 14.9, 16.4, and 17.9 hours.
\label{fig:maps_sgrastar}}
\end{figure*}

Figure~\ref{fig:total_map} shows the contour maps for both Stokes $I$ and $V$ for the whole track of March 31, 2007. In this case, the measured CP flux density
is $-41\pm3$~mJy (in the standard IAU sign convention, where a negative sign indicates LCP), corresponding to $\sim-1.2\%$ of the total flux density. The background rms noise of the image is $3.1$~mJy~beam$^{-1}$, consistent with the statistical error of  $\pm3$~mJy obtained from the visibility fit. Figure~\ref{fig:pol_data_d} shows the compilation of published CP observations from the lower panel of Figure~\ref{fig:pol_data} along with our new measurements. Although there are many unobserved frequencies, the polarization appears to increase monotonically with frequency while retaining the same handedness throughout. The CP spectrum scales approximately as $\nu^{0.35\pm0.03}$ across the range of frequencies with detections. 
Subdividing the track into four segments, (see Figure~\ref{fig:maps_sgrastar} and Table~\ref{tab:offsets}), we see no statistically significant variability in the CP, with the largest change corresponding to a 2-$\sigma$ difference. We limit fractional variation in the CP, i.e., $\Delta V/V$, at the 2-$\sigma$ level of significance to 40\%, which is comparable to the fractional changes in CP at 1$-$15~GHz measured by \citet{bow02}. There is a tantalizing correlation at the 2-$\sigma$ level of significance between the circularly and linearly polarized flux densities. However, with only four data points, so significant conclusion can be drawn. 

Figure~\ref{fig:total_map} shows a slight offset between the peak positions of $V$ and $I$. This offset, 0.18~arcseconds, is consistent with what is expected for an SNR of $\sim$14. The offset is observed in all of the Sgr~A* polarimetric tracks at all frequencies, and the orientation and angular amplitude of the offset vary with time within a single track. This shift between the peaks of the Stokes $V$ and Stokes $I$ images suggests imperfection in our CP calibration procedure. The offset is barely noticeable for maps derived from complete tracks (e.g., Figure~\ref{fig:total_map}), but it becomes significantly larger when the track is split into consecutive time intervals (see Figure~\ref{fig:maps_sgrastar} and Table~\ref{tab:offsets}). The fact that the offset grows when reducing the SNR suggests that the wandering of the Stokes $V$ map around the Stokes $I$ may be in part due to a thermal noise effect. However, extensive tests (Section~\ref{sec:errors}) suggest that the offset is not dominated by noise.

\input{table_offsets}   

Table~\ref{tab:quasars_pol} shows the polarization flux densities for all the test quasars observed on March 31, 2007. All point-source flux densities are obtained from the visibility fitting process. Quasars 3C273, 3C279, 3C286, 1337-129, and 1733-130 show CP fractions consistent with zero in both sidebands. The deviations from zero are of the same sign for all objects, which suggests a systematic error that would fractionally apply to Sgr~A* as well. However, the 0.1\% weighted-average magnitude (right CP) of this systematic error is negligibly small, an order of magnitude smaller than the observed CP in Sgr~A*.

The circular polarization of quasar 1924-292 (last line of Table~\ref{tab:quasars_pol}) was also measured in a separate track on May 30, 2008. The $V$ flux determined by a point-source fit to the visibilities, constrained to lie at the pointing center, is consistent with zero, as for the other sources. However, within a region comparable in size to the synthesized beam, the Stokes $V$ signal varies between $+5\sigma$ and $-4\sigma$, as shown in Figure~\ref{fig:1924_map}. This antisymmetric pattern and other calibration effects are discussed in detail in Section~\ref{sec:1924}. 

\input{tab_quasars_pol}     
%


\section{Error Inspection and Polarization Tests}\label{sec:errors}

The consistency of the measurements in Table~\ref{tab:sgra_pol} over a period of three years and at multiple frequencies suggests a robust detection of CP. The most sensitive of the detections has a significance of greater than 10$\sigma$. Because our work presents the first measurements of CP with the SMA, we feel it is important to describe our calibration methods and system checks in detail.


\subsection{Polarization Calibration}\label{sec:gains}

In an ideal interferometer with circularly polarized feeds, CP in a target source appears as a difference between the parallel-hand visibilities. Complications may arise due to imperfection in the interferometer response and calibration. The four polarized visibilities in a circularly polarized interferometer can be written \citep{TMS}:
\begin{equation}\label{eq:visibilities_rr}
\begin{array}{l}
\mathcal{V}_{RR}=g_{Ra}g^*_{Rb} \left[\left(\mathcal{V}_I+\mathcal{V}_V\right)+d_{Ra}d^*_{Rb}\left(\mathcal{V}_I-\mathcal{V}_V\right)\right. \\
\nonumber +\left.d_{Ra}\left(\mathcal{V}_Q-i\mathcal{V}_U\right)e^{2i\phi}+d^*_{Rb}\left(\mathcal{V}_Q+i\mathcal{V}_U\right)e^{-2i\phi}\right]\\
\\
\mathcal{V}_{RL}= g_{Ra}g^*_{Lb}  \left[\left(\mathcal{V}_Q+i\mathcal{V}_U\right)e^{-2i\phi}+d_{Ra}\left(\mathcal{V}_I-\mathcal{V}_V\right)\right.\\
\nonumber \left.
-d^*_{Lb}\left(\mathcal{V}_I+\mathcal{V}_V\right)
-d_{Ra}d^*_{Lb}\left(\mathcal{V}_Q-i\mathcal{V}_U\right)e^{2i\phi}\right]\\
\\
\mathcal{V}_{LR}= g_{La}g^*_{Rb}  \left[\left(\mathcal{V}_Q-i\mathcal{V}_U\right)e^{2i\phi}-d_{La}\left(\mathcal{V}_I+\mathcal{V}_V\right)\right.\\
\left.
+d^*_{Rb}\left(\mathcal{V}_I-\mathcal{V}_V\right)
-d_{La}d^*_{Rb}\left(\mathcal{V}_Q+i\mathcal{V}_U\right)e^{-2i\phi}\right]\\
\\
\mathcal{V}_{LL}=g_{La}g^*_{Lb}  \left[\left(\mathcal{V}_I-\mathcal{V}_V\right)+d_{La}d^*_{Lb}\left(\mathcal{V}_I+\mathcal{V}_V\right)\right.\\
\nonumber \left.-d_{La}\left(\mathcal{V}_Q+i\mathcal{V}_U\right)e^{-2i\phi}-d^*_{Lb}\left(\mathcal{V}_Q-i\mathcal{V}_U\right)e^{2i\phi}\right] ~~,
\end{array}
\end{equation}
where $g_{Ra}$ and $g_{La}$ ($d_{Ra}$ and $d_{La}$) are the right and left circular feed gains (polarization leakage terms) of antenna $a$, and $\phi$ is the parallactic angle of the feed.
Solving the linear system of equations above for the Stokes $I$ and the Stokes $V$ visibilities, respectively, gives
\begin{equation}\label{eq:visibilities_i}
\begin{array}{l}
\mathcal{V}_I=\cfrac{\mathcal{V}_{RR}}{2g_{Ra}g_{Rb}^*}\cfrac{(1+d_{La}d_{Lb}^*)}{(1+d_{Ra}d_{La})(1+d_{Rb}^*d_{Lb}^*)}\\
+\cfrac{\mathcal{V}_{LL}}{2g_{La}g_{Lb}^*}\cfrac{(1+d_{Ra}d_{Rb}^*)}{(1+d_{Ra}d_{La})(1+d_{Rb}^*d_{Lb}^*)}\\
+\cfrac{\mathcal{V}_{RL}}{2g_{Ra}g_{Lb}^*}\cfrac{(d_{La}-d_{Rb}^*)}{(1+d_{Ra}d_{La})(1+d_{Rb}^*d_{Lb}^*)}\\
+\cfrac{\mathcal{V}_{LR}}{2g_{La}g_{Rb}^*}\cfrac{(d_{Lb}^*-d_{Ra})}{(1+d_{Ra}d_{La})(1+d_{Rb}^*d_{Lb}^*)}
\end{array}
\end{equation}
and
\begin{equation}\label{eq:visibilities_v}
\begin{array}{l}
\mathcal{V}_V=\cfrac{\mathcal{V}_{RR}}{2g_{Ra}g_{Rb}^*}\cfrac{(1-d_{La}d_{Lb}^*)}{(1+d_{Ra}d_{La})(1+d_{Rb}^*d_{Lb}^*)}\\
-\cfrac{\mathcal{V}_{LL}}{2g_{La}g_{Lb}^*}\cfrac{(1-d_{Ra}d_{Rb}^*)}{(1+d_{Ra}d_{La})(1+d_{Rb}^*d_{Lb}^*)}\\
-\cfrac{\mathcal{V}_{RL}}{2g_{Ra}g_{Lb}^*}\cfrac{(d_{La}+d_{Rb}^*)}{(1+d_{Ra}d_{La})(1+d_{Rb}^*d_{Lb}^*)}\\
-\cfrac{\mathcal{V}_{LR}}{2g_{Rb}^*g_{La}}\cfrac{(d_{Ra}+d_{Lb}^*)}{(1+d_{Ra}d_{La})(1+d_{Rb}^*d_{Lb}^*)}~~.
\end{array}
\end{equation}
To first order in the leakages $d$, and ignoring terms proportional to $\mathcal{V}_{LR}\,d$ and $\mathcal{V}_{RL}\,d$ (small LP as well as small leakages), we have
\begin{equation}\label{eq:i_vis}
\mathcal{V}_{I}\simeq\frac{1}{2}\left\{\frac{}{}\mathcal{V}_{RR}/(g_{Ra}g_{Rb}^*)+\mathcal{V}_{LL}/(g_{La}g_{Lb}^*)\right\}
\end{equation}
and
\begin{equation}\label{eq:v_vis}
\mathcal{V}_{V}\simeq\frac{1}{2}\left\{\frac{}{}\mathcal{V}_{RR}/(g_{Ra}g_{Rb}^*)-\mathcal{V}_{LL}/(g_{La}g_{Lb}^*)\right\}~~,
\end{equation}
and thus Stokes $I$ and $V$ are independent of the leakages to first order.

The MIRIAD reduction package \citep{sau95} uses these first-order equations when solving for the polarized leakages, ignoring second-order terms in the leakages $d$ and LP
fraction. These terms contribute a systematic error in Stokes $V$ of the form $Id^2$ and $md$, for LP fraction $m$.
For the track of March 31, 2007, all leakage terms are of order $\sim10^{-2}$ (both real and
imaginary parts), introducing a fractional contribution from Stokes $I$ of $\sim10^{-4}$. Likewise, the terms of order $md$ are of order $10^{-3}$ and are unable to explain the observed CP fraction of $10^{-2}$. Another uncertainty arises from the complex terms dependent on the parallactic angle in the form of a phase term in the full gain equations. A systematic phase effect could, in principle, explain offsets in the image plane; however,
the low flux density contribution from these offset components ($\sim md$) makes this possibility difficult to reconcile with the evident displacement of the entire point-source flux density in Figure~\ref{fig:maps_sgrastar}.


\begin{figure}[ht!]  
\epsscale{1.0}
\plotone{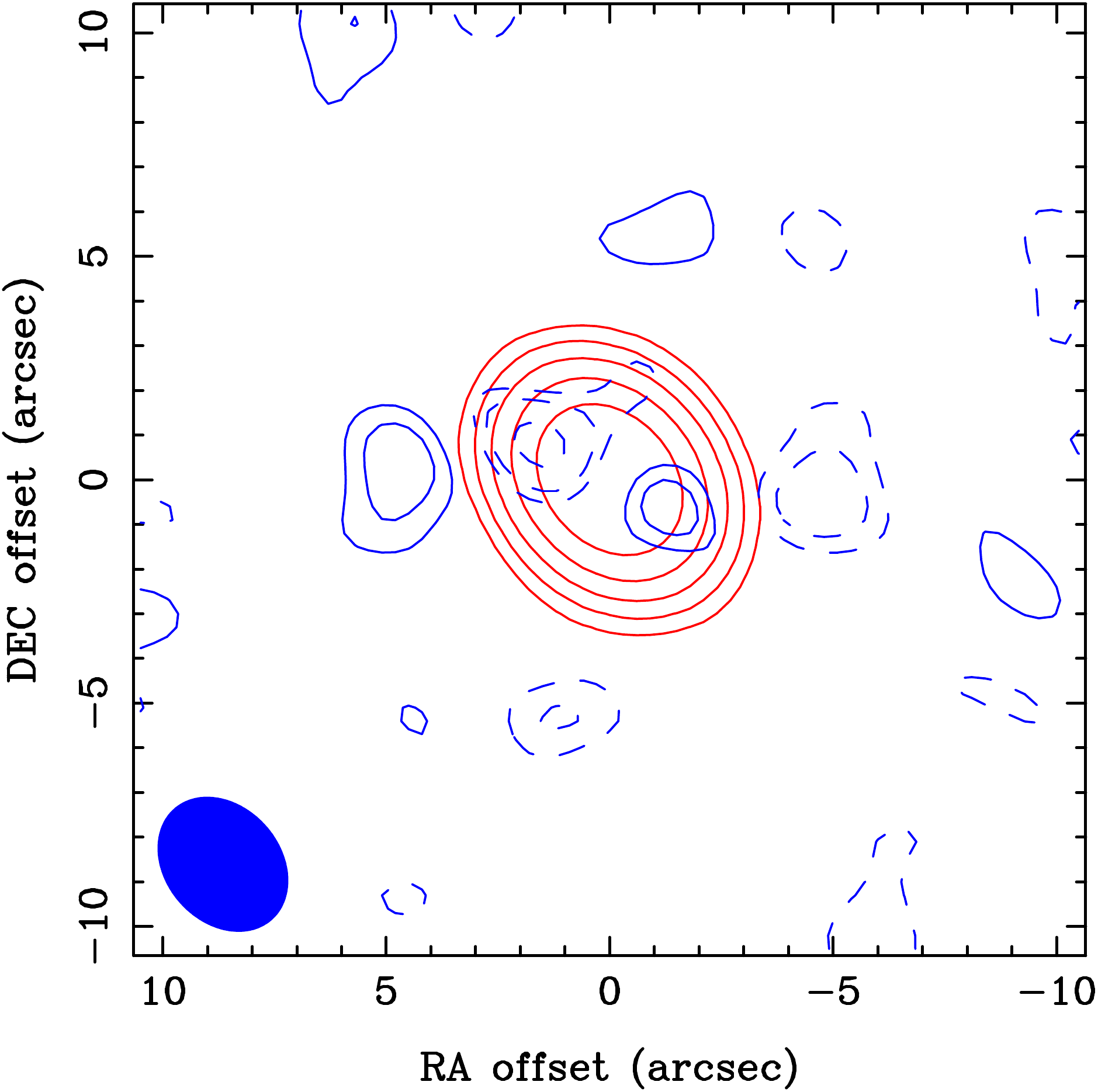}
\caption{Total and CP emission of 1924-292. Color coding is identical to Figure~\ref{fig:total_map}. Stokes
$I$ contour spacings are $12,25,50,100,200\,\times\sigma_I$, where $\sigma_I=14.7$ mJy/beam is the rms noise in the $I$ map. The contour levels for the Stokes $V$ map are 
$-5,-3,-2,2,3\times\sigma_V$ where $\sigma_V=5.85$ mJy/beam is the rms noise in the $V$ map. Within the angular extent of the Stokes $I$ point source, two distinct Stokes $V$ are identifiable. The negative peak corresponds to approximately $-30$ mJy/beam (i.e., $-0.4\%$ of the Stokes $I$ peak), and the positive one corresponds to approximately $+18$ mJy/beam (i.e., $0.25\%$ of the Stokes $I$ peak).
\label{fig:1924_map}}
\end{figure}

\subsection{Gain Calibration}\label{sec:Agains}
Errors in the relative calibration of the $L$ and $R$ gains of the antennas affect the precise cancellation of terms in eq.~\ref{eq:v_vis} and are the most natural way to generate false CP. The time variation of the antenna gains is monitored using 1733-130, which, at a flux density of $\sim$1~Jy, is much weaker than \sgr. In order to maximize the signal-to-noise of the gain measurements, we choose to average the amplitude of the $LL$ and $RR$ visibilities together and determine a single gain curve per antenna rather than separately measuring $g_L$ and $g_R$. If there is an imbalance in the response of the two ``feeds,'' which are really the same receiver looking through the same quarter-wave plate, this will introduce false CP.

The first test of our calibration method is to examine the calibrated 1733-130 data for CP. In Table~\ref{tab:quasars_pol} we report the polarization measured in 1733-130 after applying the same gains that are used with the \sgr\ data. Although we make no attempt to correct the difference in the $L$ and $R$ gains in our calibration procedure, we end up with little polarization in this source ($<$0.1\%), with a fractional $V$ sensitivity of $\sim5\times10^{-4}$. This suggests a systematic polarization limit of $10^{-3}$ for this calibration method. Performing the calibration with separate gains for $L$ and $R$ yields nearly identical results for the polarization fraction of \sgr\ but with slightly higher noise.

\begin{figure*}[t]  
\epsscale{0.35}
\begin{center}
\begin{tabular}[c]{cc}
  \begin{tabular}[c]{c}
\plotone{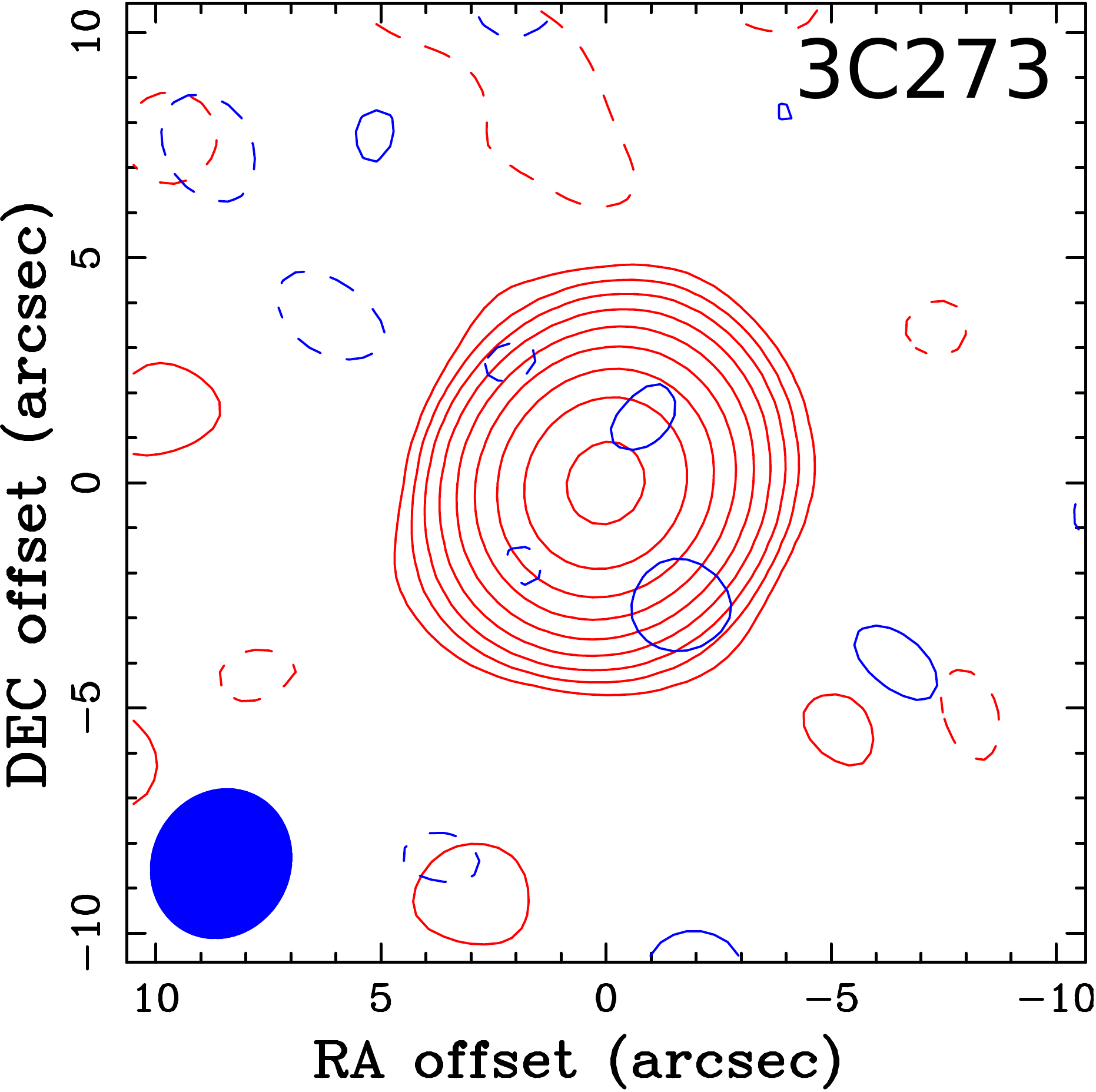}
  \end{tabular} &
  \begin{tabular}[c]{c}
 \plotone{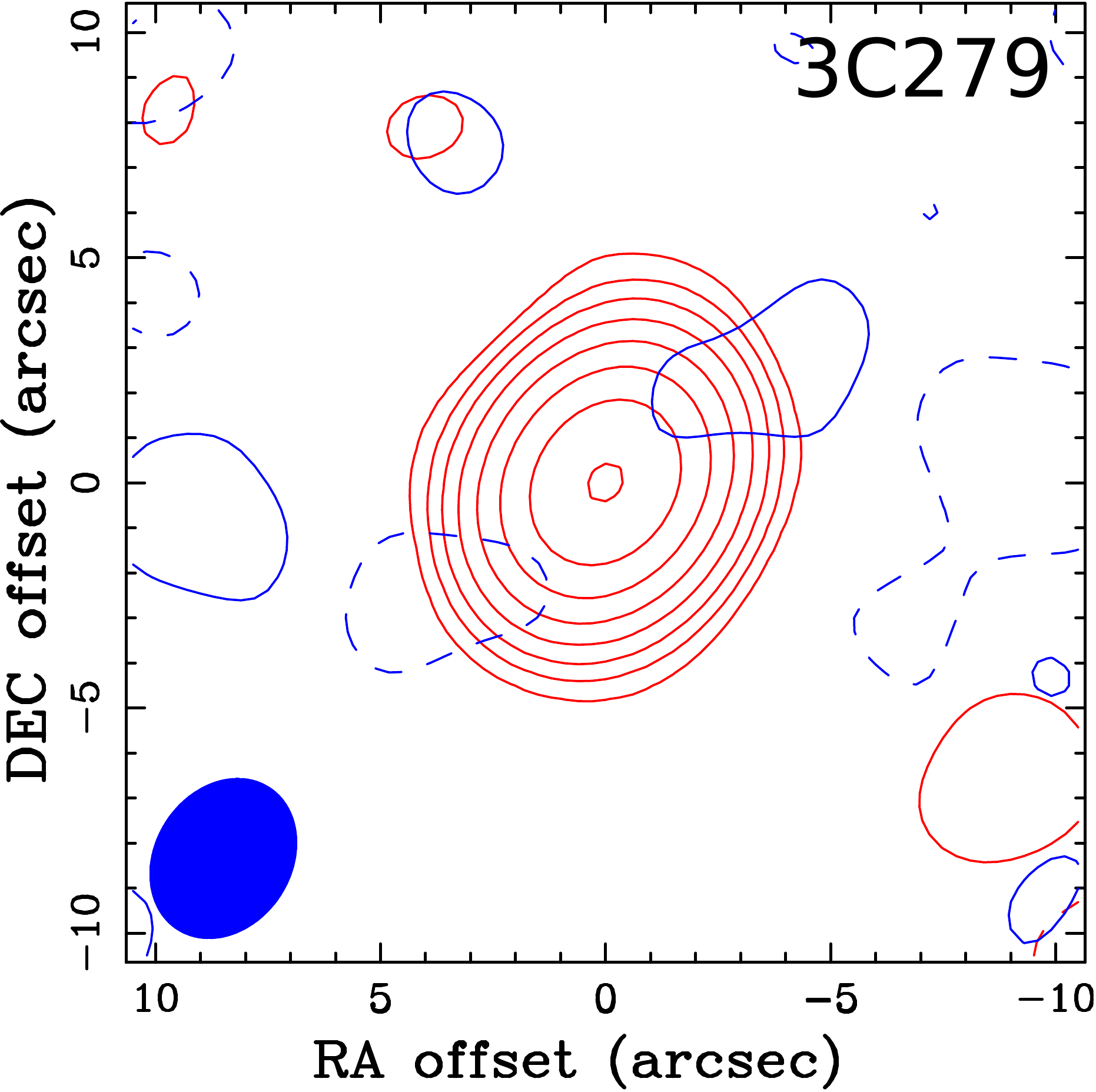}
   \end{tabular} \\
  \begin{tabular}[c]{c}
\plotone{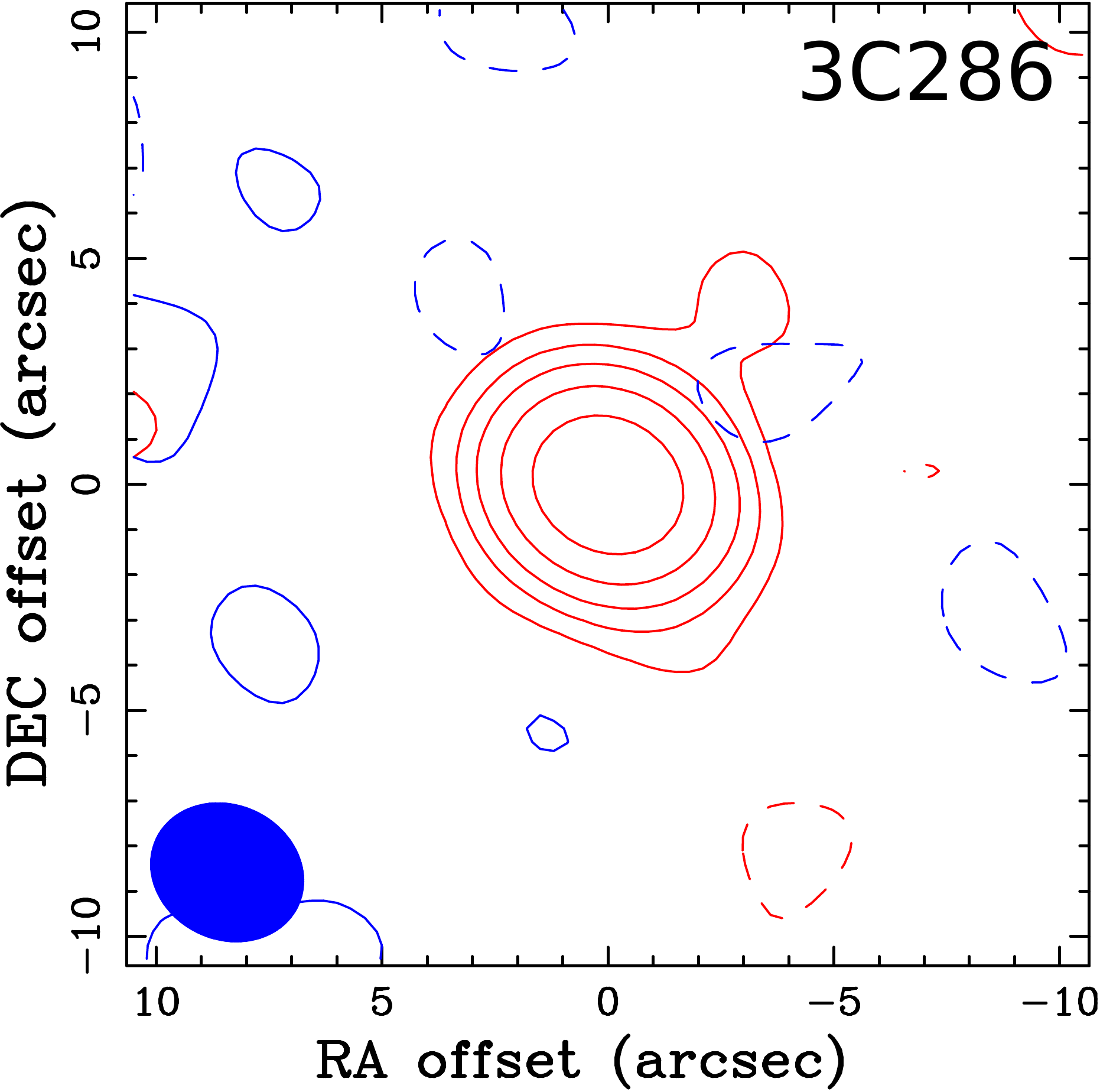}
  \end{tabular} &
  \begin{tabular}[c]{c}
\plotone{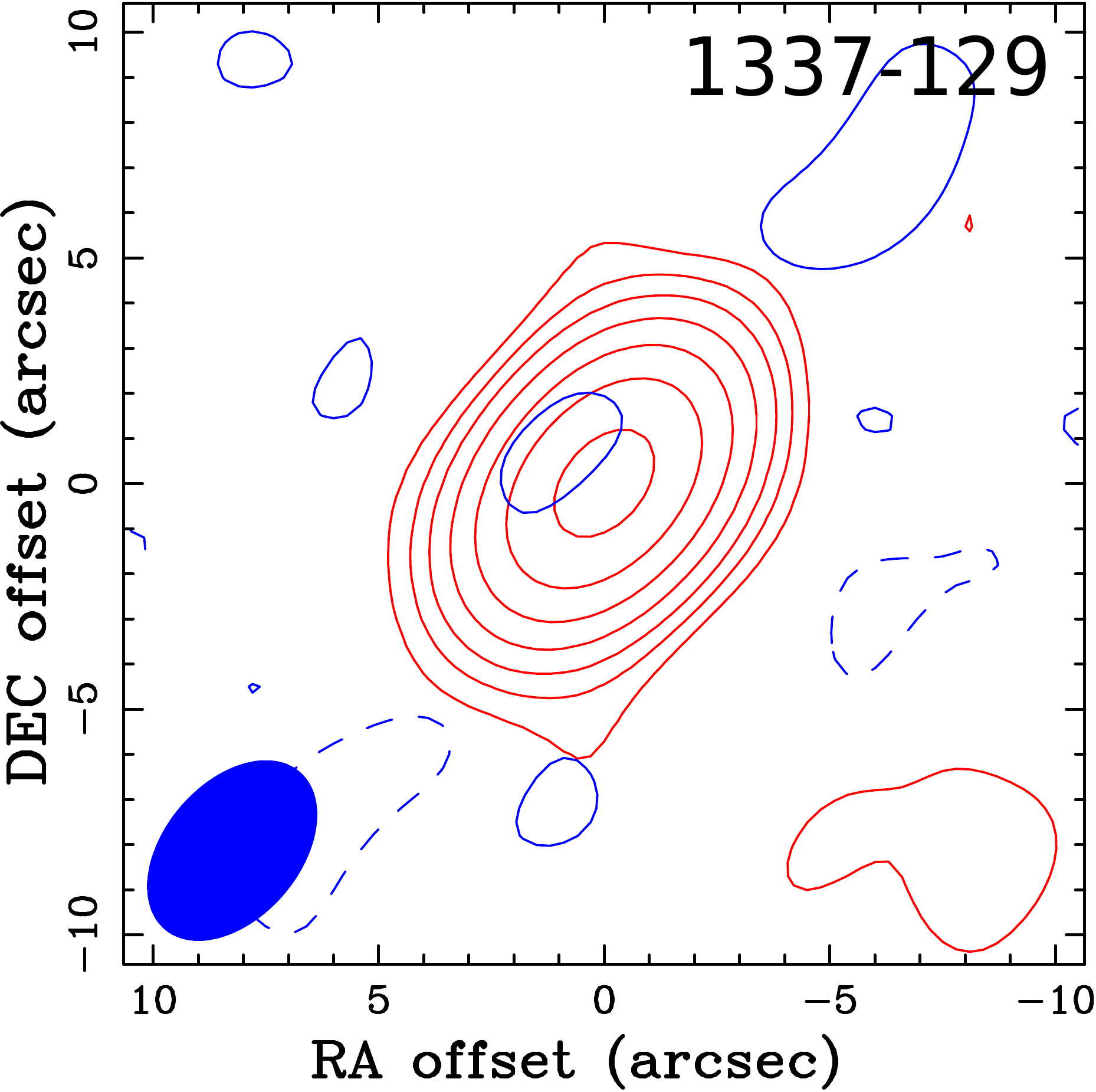}
  \end{tabular} \\
\end{tabular}
\end{center}
\caption{Quasars 3C273, 3C279, 3C286, and 1337-129 observed during the track of March 31, 2007, after gain calibration using 3C273 as calibrator. Each image shows contour plots in Stokes $I$ and
$V$ in the same way as Figure~\ref{fig:total_map}. For each panel, the contours in Stokes $V$ are --2 and 2 times the rms noise of the image, and all four of them have maxima and minima between $-3\sigma_V$ and $+3\sigma_V$. The intensity contours in Stokes $I$ are  $(-2,2,5,10,20,40,80,160,320,640)\times\sigma_I$
for 3C273 and 3C279. For 3C286 and 1337-129, the contour levels are $(-2,2,5,10,20,40)$ and $(-2,2,5,10,20,40,80,160)\times\sigma_I$, respectively.
\label{fig:quasarsmap}}
\end{figure*}

A second test of this calibration method is to examine the $V/I$ measured for other quasars in the track. We expect no significant CP in quasars at these wavelengths. Figure~\ref{fig:quasarsmap} and Table~\ref{tab:quasars_pol} show that no CP is found in the quasars observed in the March 31, 2007, track. Fractional polarization limits are comparable to those obtained on 1733-130, except for 3C286, which is significantly fainter than the others and thus has a higher level of noise in $V/I$.

\subsection{Circular Polarization of 1924-292}\label{sec:1924}
To guard against poorly understood instrumental effects that preferentially affect, for example, sources at low elevation or those with poor parallactic angle coverage, we undertook a supplementary test using the quasar 1924-292. The declination of this source is within half a degree of that of \sgr, ensuring that it follows the same path on the sky. The observation scheme was designed to precisely mimic that of the primary track on \sgr, with similar hour angle coverage and temporal sampling, identical polarization modulation, and  a gain calibrator (1922-201) at a similar distance from the target ($9^\circ$ north compared to $16^\circ$ for 1733-130 and \sgr). Because 1924-292 was brighter at the time of these observations, 7~Jy, than \sgr\ was during the other observations (3--4~Jy), and because of the long integration time in the track, this test also provides a more sensitive limit on instrumental CP than the that provided by observations of quasars on March 31, 2007. 

The results of this test are shown in Figure~\ref{fig:1924_map} and the last line of Table~\ref{tab:quasars_pol}. Fitting the visibilities to a point source, as was done for other quasars and \sgr, we find $V/I = 1-2\times10^{-4}$, with an uncertainty of $10^{-3}$. This is a factor of more than 10 lower than the CP fraction observed in \sgr, including that measured in the same track in a short observation (last line, Table~\ref{tab:sgra_pol}). That the polarization of 1924-292 was not found at the same time that the previously observed CP level in \sgr\ was reconfirmed enhances our confidence in our result. 

Figure~\ref{fig:1924_map} does show significant CP in an antisymmetric pattern around the pointing center, with a peak $V/I$ values of +2.5 and $-4.2\times10^{-3}$. The negative peak is approximately five times the image rms. The antisymmetric pattern is clearly distinguishable from the offset CP detections in \sgr, which do not show peaks of opposite sign one resolution element away from the main peak. We conclude that this is a signature of an unrelated calibration problem. This calibration could mask a peak in the $V$ map that is offset from the pointing center as seen in \sgr\ in Figure~\ref{fig:maps_sgrastar}, with the allowed amplitude being approximately the difference in the positive and negative peaks, or $-1.7\times10^{-3}$. We take this as a limit on our uncorrected instrumental polarization, $V/I<2\times10^{-3}$. 

The LP of 1924-292 was measured to be 10.8\% during this track, larger than was observed in the \sgr\ tracks in this work. The increased LP makes this source slightly more susceptible to errors from ignored second-order terms from feed imperfections (Section~\ref{sec:gains}). Nevertheless, no significant CP is detected.

\subsection{Other Possible Sources of Error}
In the tests discussed thus far, we failed to find CP in any source other than \sgr, placing tight limits on the possible instrumental contribution to the observed signal. The previous section describes the most stringent test, which mimics everything about the \sgr\ observation and nevertheless finds insignificant CP in a bright test source. 

A possible distinction between the test sources and \sgr\ is the presence of extended emission in our Galactic Center. The test sources are nearly pointlike (with some low-level emission from jet knots in the case of 3C279), while \sgr\ is embedded in dust and free-free emission. To avoid contamination from larger-scale emission, we exclude baselines shorter than 20~k$\lambda$ from our visibility fits and find that the same exclusion in other sources does not change our results. We are not aware of a mechanism by which the extended emission could introduce false CP, but we have also attempted to verify that no such effect is present. We examined a data set from July 17, 2006, which measures $uv$ spacings of 27--390~k$\lambda$ instead of the 5--50~k$\lambda$ that is typical of the other tracks. This track should be more immune to the larger scale emission, which is highly suppressed by the shorter fringe spacing. We see no significant difference in Table~\ref{tab:sgra_pol} between this track and the others.

\section{Astrophysical Sources of Circular Polarization}\label{sec:cp_sources}
The polarization state of radiation emerging from an astrophysical source is governed by radiative transfer equations that, in their most complete form, incorporate emission and absorption in each Stokes parameter. This polarized radiative transfer equation must also include mixing between Stokes parameters that takes place in birefringent media, such as a synchrotron plasma. The magnetic field that is fundamental to the synchrotron emission and self-absorption is also a source of birefringence. Variations in the magnetic field along the line of sight, and the change in the coupling between Stokes parameters with plasma temperature, complicate the picture further. A fully consistent treatment requires addressing all these effects simultaneously.

\subsection{Polarized Transfer in a Homogeneous Magnetic Field}
\sgr, a stratified, self-absorbed synchrotron source \citep[see][for the basics of these models]{bru76} with frequency- and time-variable polarization, clearly requires a sophisticated radiative transfer model.  The Faraday rotation inferred from submillimeter observations \citep{marr07} has been assumed to be separate from the submillimeter emission regions, although it may instead
occur deep inside the source \citep[sometimes referred to as {\it internal Faraday rotation}, e.g.,][]{war03}. The production of CP may similarly be a signature of conversion processes within the emission region. The complete PRT equation is \citep[e.g.,][]{saz69}
\begin{eqnarray}\label{eq:pol_transfer_eq}  
\cfrac{d}{dl}\left(\begin{array}{c}I\\ Q\\ U\\ V\end{array}\right)=
\left(\begin{array}{c}\eta_I\\ \eta_Q\\ 0 \\ \eta_V \end{array}\right)
 +
\left(\begin{array}{cccc}
-\kappa_I &-\kappa_Q&0&-\kappa_V\\
-\kappa_Q &-\kappa_I&-\kappa_V^*&0\\
0&+\kappa_V^* &-\kappa_I&-\kappa_Q^*\\
-\kappa_V&0&+\kappa_Q^*&-\kappa_I\\
\end{array}\right)\left(\begin{array}{c}I\\ Q\\ U\\ V\end{array}\right)~~,
\end{eqnarray}
where the cartesian axes in the plane of the sky ($\hat{\mathbf{e}}_1$ and $\hat{\mathbf{e}}_2$)---which determine the precise definition of Stokes $Q$ and $U$---are oriented such that $\hat{\mathbf{e}}_1$ is
aligned with the local direction of the magnetic field. The absence of transport coefficients related to Stokes $U$ in eq.~\ref{eq:pol_transfer_eq} is a direct consequence of this convenient choice of coordinates, and thus it is a valid expression only locally. 

Eq.~\ref{eq:pol_transfer_eq} is the vectorial generalization of the standard radiative transfer equation $dI_\nu/dl=j_\nu-\alpha_\nu I_\nu$.  The $\eta_A$ (with $A=I,Q,U,V$) are the emission coefficients for each Stokes parameter.
The absorption coefficients 
$\kappa_A$ (one for each Stokes parameter) are related to the $\eta_A$ through detailed balance. The antisymmetric portion of the matrix comprises the Faraday transport coefficients $\kappa_A^*$.
They couple $Q$, $U$, and $V$, permitting exchanges of LP and CP along the path of radiation. This rotation in three-dimensional polarization space ($Q,U,V$) has been referred to as the {\it generalized Faraday rotation} effect \citep{mel97}. The general Faraday effect reduces to pure Faraday rotation (i.e., rotation in the $Q$--$U$ plane) in the nonrelativistic limit  and to pure Faraday conversion (i.e., rotation in a plane perpendicular to the $Q$--$U$ plane) in the ultrarelativistic limit but, in general, is a combination of both.

The intrinsic emission term in 
eq.~\ref{eq:pol_transfer_eq} ($\eta_V$) demonstrates the first potential source of CP. The intrinsic emission in Stokes $V$ is weak: $\eta_V$ is of order 1/$\gamma$ smaller than the other components and is generally ignored in synchrotron emitting plasma. On the other hand,
predominantly CP emission can be expected in a cold cyclotron emitting plasma \citep[e.g.,][]{lan75}, in which the relativistic approximations intrinsic to synchrotron spectra \citep[e.g., beaming, see][]{ryb79} are not valid \citep[see also][]{mah96}. The submillimeter emission from \sgr\ is expected to arise at very small radius where the plasma is hot, and therefore emission from cold electrons is not expected to be relevant.

The birefringence of the plasma may also transform LP to CP, as noted above, but this ``Faraday conversion'' is weak in a cold plasma. 
In this limit, the response of the medium depends exclusively on the plasma frequency, $\nu_p=(n_e e^2)^{1/2}/(\pi m_e)^{1/2}$, and the cyclotron frequency, $\nu_B=eB/(2\pi m_ec)$. In the high frequency regime (i.e., $\nu_B/\nu\ll1$, applicable to the submillimeter emission considered here given plausible magnetic field strengths), the rotation and conversion coefficients are \citep{mel05,swa89}
\begin{eqnarray}
\label{eq:kappaV} {\kappa_V^*}_\mathrm{cold}&=&(2\pi/c)(\nu_p^2\nu_B/\nu^2)\cos\theta \\
\label{eq:kappaQ} {\kappa_Q^*}_\mathrm{cold}&=&-(\pi/c)(\nu_p^2\nu_B^2/\nu^3)\sin^2\theta~,
\end{eqnarray}
where $\theta$ is the angle between the magnetic field vector and the line of sight. The ratio of these equations shows Faraday conversion ($\kappa_V^*$) to be a factor of $\nu_B/\nu$ ($\sim10^{-3}$ for our data) smaller than rotation, inconsistent with our observations. Equations for higher temperature plasmas are presented in section~\ref{sec:FC_hot}.

\subsection{Magnetic Field Changes}

\citet{hom09} list a series of scenarios in which CP can be produced in quasars. By incorporating line-of-sight changes in the magnetic field geometry, they offer possibilities not present in the homogenous-field radiative transfer equation
(eq.~\ref{eq:pol_transfer_eq}). In a uniform field, Stokes $Q$ needs to be converted into Stokes $U$ before Stokes $U$ can be converted into Stokes $V$. Changes in the field orientation along the line of sight break a fundamental assumption of eq.~\ref{eq:pol_transfer_eq}:
the azimuthal angle ($\phi$) between the field direction in the plane of the sky and the coordinate axis $\hat{\mathbf{e}}_1$ was defined to be zero. Changes in field direction will introduce a coupling of $Q$ and $V$ across the domains of magnetic orientation.
The magnitude of the transfer coefficients also depends on the angle $\theta$, introducing a small additional change in coupling. These field orientation changes, whether stochastic \citep{rus02} or ordered (as in the helical fields of \citealt{beck03}) can also create significant CP. Both of these mechanisms are capable of producing the CP sign coherence observed in \sgr, although a small net bias to the field direction is important for the stochastic field.


\subsection{Self-Absorbed Faraday Conversion}\label{sec:selfabsorbed}
The wavelength-dependent size of \sgr\ indicates the importance of synchrotron self-absorption in the observed SED. The source size and brightness are determined by the variation of density, magnetic field, and electron energy distribution with radius/distance, although there are unbreakable degeneracies between these quantities even assuming power laws in radius without additional information or assumptions. A specific choice of jet structure can determine the structure in density/field/energy and match both size and flux \citep[e.g.,][]{mar07} and even variability timing \citep{fal09}. Similarly, the assumption of, e.g., an equipartition magnetic field can specify the structure for an accretion flow while matching observations.

Polarimetric observations add new information regarding the source structure, potentially reducing the need for unverifiable assumptions. The structure of \sgr\ is likely to be more complex than is captured by power-law models; this is particularly true at submillimeter wavelengths where the emission originates near the black hole. Complex magneto-hydrodynamic simulations and general relativistic radiative transfer \citep[e.g.,][]{hua08,mos09,shch10b,shch11} will more faithfully represent the true source properties near 345~GHz than few-parameter analytic models. Nevertheless, the smoothness of the average CP spectrum, the exclusive left-handedness of the polarization from 1.4 to 345 GHz, and the slow monotonic variation of the fractional polarization suggest that much insight can be gained from a simple conceptual model incorporating the ideas of self-absorbed synchrotron sources \citep[e.g.,][]{bru76}.

In the case of self-absorbed synchrotron sources, the specific intensity near frequency $\nu$ is dominated by  the emission of electrons within a narrow energy range around $\gamma_\mathrm{rad}$. An electron with Lorentz
factor  $\gamma_\mathrm{rad}$ emits most of its radiative power at a frequency $\nu_c\sim\gamma_\mathrm{rad}^2eB\sin\theta/(2\pi m_ec)=\gamma_\mathrm{rad}^2\nu_{B\perp}$. The brightness temperature associated with the emitting electrons is related to this effective Lorentz factor by \citep[e.g.,][]{pac70,ryb79} 
$ k_B T_b=\alpha(p) \gamma_\mathrm{rad}m_ec^2$,
 where $\alpha(p)$ is a coefficient of order unity. 
  The brightness temperature can
 be obtained directly from observations: $T_b=1.22\times10^{12}S_\nu\nu^{-2}\theta_s^{-2}(\nu){~\rm K}$; where $S_\nu$ is the observed flux density in Jy, $\theta_s$ is the source angular radius in milliarcseconds, and $\nu$ is in GHz. Assuming power-law dependencies of the form $S_\nu\propto\nu^{-m}$ and $\theta_s(\nu)\propto\nu^{-n}$, we obtain $\gamma_\mathrm{rad}\propto\nu^{-m-2+2n}$ \citep[see, for example,][]{loe07}.

\subsection{Self-Absorbed, Stratified Sources}

In what follows, our main approximation is to assume that at each frequency, the synchrotron-emitting electrons, with energies $\gamma_\mathrm{rad}$, are also responsible for Faraday conversion. We also assume that Faraday rotation and conversion can be considered to act in an alternating manner rather than attempting to treat the full, complicated PRT equation. The radiative transfer near the $\tau=1$ surface is particularly complicated because all matrix elements can be important. By assuming sequential effects, we can operator-split the differential equation. In this case, and for short propagation distances, the production of Stokes $V$ is simply an angle of rotation in the Poincair\'e sphere corresponding to the cumulative phase shift between the polarization modes of the plasma near radius $r_0$ \citep{kenn98}. This can be calculated as an integral of $\kappa_Q^*$,
\begin{equation}\label{eq:conversion}
\Delta\psi(r_0,\nu)=\int_{r_0}^\infty \cfrac{\pi}{c}\cfrac{\nu_p^2\nu_B^2}{\nu^3}\sin^2\theta~\Theta_\mathrm{FC}~dr~~,
\end{equation}
where we have corrected the cold-plasma Faraday conversion rate (eq.~\ref{eq:kappaQ}) by a factor $\Theta_\mathrm{FC}$ (see Section~\ref{sec:FC_hot} below) that takes into account the effect of a power-law distribution of relativistic electrons
on Faraday conversion.

Following the treatment of Faraday conversion presented by \citet{kenn98}, we take the finite-temperature correction to be  proportional to the local mean Lorentz factor of the electron distribution, and we further approximate this with our value of $\gamma_\mathrm{rad}$ \citep[in the same approximation $ \Theta_\mathrm{FR}\sim \log\langle\gamma\rangle/\langle\gamma\rangle$; e.g.,][]{saz69,qua00a},
\begin{equation}
 \Theta_\mathrm{FC}(\nu,T_e,\theta,B)\sim\langle\gamma(r)\rangle\sim \gamma_\mathrm{rad}~~.
\end{equation}
Assuming conversion in a narrow range in $r$, we write
\begin{equation}\label{eq:conversion2}
\begin{split}
\Delta\psi(r_0,\nu)&\propto \int_{r_0}^\infty n_e(r)B(r)^2\sin^2\theta~ \gamma_\mathrm{rad}(r) \nu^{-3} dr\\
&\sim n_e(r_0)B(r_0)^2~ \gamma_\mathrm{rad}(r_0)\nu^{-3}r_0~~.
\end{split}
\end{equation}
We can obtain $\Delta \psi$ as a function of frequency  by using the fact that for each observed frequency $\nu_0$, there is a corresponding radius $r_0$. If we assume, as well, that
the electron density and magnetic field have radial profiles of power-law form, $n_e\sim r^{-\beta}$ and $B\sim r^{-\alpha}$,
the Faraday conversion phase shift is
\begin{equation}\label{eq:conversion3}
\Delta\psi(r_0,\nu_0)\sim r_0^{-\beta-2\alpha+1}\nu_0^{-m+2n-5}\sim \nu^{\beta n+2\alpha n -m+n-5}~~.
\end{equation}

Power-law relations of the form $S_\nu\propto\nu^{-m}$---for the flux density---and $\theta_s(\nu)\propto\nu^{-n}$---for the source angular size---are a natural consequence of stratified synchrotron sources \citep{bru76}. In this simple model, the size-frequency relation is
obtained as follows. First, we take the synchrotron optical depth associated with a narrow shell of radius $r$ \citep{ryb79}: $\tau_\nu(r)=\alpha_\nu \Delta r\sim n_e(r)B^{(p+2)/2}(r)\nu^{-(p+4)/2}r\sim r^{-\beta-\alpha(p+2)/2+1}\nu^{-(p+4)/2}$, where $p$ is the electron power-law index. Hence, for a given frequency $\nu_0$, the $\tau = 1$-surface occurs at $ r_0^{-\beta-\alpha(p+2)/2+1}\nu_0^{-(p+4)/2}\sim1$ or $r_0\sim\nu_0^{-(p+2)/(2\beta+\alpha(p+2)-2)}$. Similarly, the flux density can be shown to be  \citep[see][for details]{bru76} $S_\nu\sim\nu^{(13-5\beta-3\alpha-2\alpha p+2p)/(2-2\beta-2\alpha-\alpha p)}$.  The resulting spectral indices are
\begin{equation}
\begin{split}
n=\frac{p+4}{2\beta+\alpha(p+2)-2}~~~~~{\rm and}~~~~~ \\
m=-\frac{13-5\beta-3\alpha-2\alpha p+2p}{2-2\beta-2\alpha-\alpha p}~~~~.
\end{split}
\end{equation}
Therefore, $\alpha$ and $\beta$ can be solved for as functions of $m$ and $n$:
\begin{equation}
\begin{split}
\beta &=\frac{-3-2m+5n-2p-mp+2np}{n}~~~~~{\rm and}~~~~~ \\
\alpha &=\frac{5+2m-4n}{n}~~~,
\end{split}
\end{equation}
which gives (dropping the subscript in $\nu_0$ for clarity)
\begin{equation}
\begin{split}
\Delta\psi\sim\nu^l,~~~~{\rm with}~~
l & =(2n-m-2)(p-1)\\ &=\frac{(p-1)(-\alpha+\beta-1)}{\alpha(p+2)+2(\beta-1)}~~~.
\end{split}
\end{equation}

The indices $\alpha$ and $\beta$ are very sensitive to $n$, which is close to 1. \citet{fal09} report a value of $n=1.3\pm0.1$, obtained from observations between 22~GHz and 230~GHz, after correcting for interstellar scattering. These results are very sensitive
to the scattering model. Another approach, used by \citet{she06}, is to minimize the dependence
on the scattering law by relying on the shortest wavelength data. For measurements at 43 and 86~GHz, their result is $n=1.09\pm0.3$. An even shallower slope can be obtained using the results from \citet{doe08}. From the VLBI visibilities, the angular size of the source, assuming a circular gaussian profile, is 37~$\mu$as, which is smaller than the last photon
orbit for a Schwarzschild black hole ($\sim52~\mu$as). If the source geometry is assumed to be an annulus \citep{doe08}, the mean diameter is 58~$\mu$as. Using this angular diameter at 230~GHz gives a value of $n\sim0.8$. In particular, for this value of $n$, with $m=-0.43$ \citep{an05}, and assuming an electron power-law index of 2.4, we obtain $\alpha\sim 1.2$ and
$\beta\sim1.4$. These values are  close to those corresponding to spherical accretion and equipartition of energy ($\beta=3/2$ and $\alpha=(\beta+1)/2=5/4$). 
Most importantly, these indices imply $l\sim0.1$, i.e., the CP production
rate is almost independent of frequency, which is consistent with  observations (ranging from 0.5$\%$ to 1.5$\%$) across a range in frequency
 of almost  two orders of magnitude. The main point here is that a mild increase of CP with frequency is possible. Although this is contrary to the decreasing 
 efficiency of Faraday conversion with frequency that is expected for a uniform medium, it is a natural consequence of stratified emission.
 
 \subsection{Magnetic Field Structure and Polarization Stability}

The amount of CP generated from LP is proportional to this phase shift, i.e., $m_c\sim m_l\,\Delta\psi$ for small $\Delta\psi$. The weak dependence of $\Delta\psi$ on frequency implies that CP production in different layers of the source behaves similarly regardless of  location respect to the SBMH.  The value of $\Delta\psi$ should remain small to avoid conversion of CP back to LP at high frequencies. Starting with the $\sim10\%$ \citep[e.g.,][]{pac70} intrinsic polarization fraction of the optically thick synchrotron, achieving CP fractions of 0.1\% to 1\% requires $\Delta\psi\sim0.01$--0.1. Such small values are far from reversing the sign of CP, consistent with the fixed sign of CP at all frequencies, as long as the orientation of the orientation magnetic field does not fluctuate randomly. A variable orientation of the magnetic field with radius---i.e., field reversals---is inconsistent with the apparent self-similarity of the CP spectrum. If this self-similarity persists for more than two orders of magnitude in frequency (1 to 345~GHz), the orientation of the magnetic field must be highly coherent over spatial scales spanning nearly two orders of magnitude (since $n\approx1$), roughly between 1$R_S$ and 100$R_S$.

In plane-of-the-sky coordinates aligned with $\mathbf{B}_\mathrm{sky}$ (total field equals plane-of-the-sky field $\mathbf{B}_\mathrm{sky}$ plus line-of-sight field $\mathbf{B}_\mathrm{los}$), eq.~\ref{eq:pol_transfer_eq} shows that for intrinsic synchrotron emission only in $I$ and $Q$, Faraday conversion proceeds as $Q\rightarrow U\rightarrow V$, i.e., conversion is driven by Faraday rotation within a narrow synchrotron shell \citep{beck03}.  The sign of $U$ generated by this mechanism depends on the
sign of the Faraday rotation coefficient $\kappa_V^*$ (eq.~\ref{eq:kappaV}), which can
change depending on the angle the magnetic field makes with respect to the line of sight $\theta$. On the other hand, the sign of the Faraday conversion coefficient $\kappa_Q^*$ (eq.~\ref{eq:kappaQ}) does not change with $\theta$. This implies that the sign of Stokes $V$ generated by conversion is exclusively determined by the sign of Stokes $U$ generated from $Q$ by local Faraday rotation, and is ultimately dependent on the local value of $\cos\theta$. Therefore, a constant sign in Stokes $V$ needs a constant orientation of  $\mathbf{B}_\mathrm{los}$. This requirement is already hinted at by the observed RM properties \citep{marr06a,marr07}.  On the other hand, the orientation of  $\mathbf{B}_\mathrm{sky}$ is free to change in time without affecting the value of $V$. This is a consequence of the sequence $Q\rightarrow U\rightarrow V$ given by eq.~\ref{eq:pol_transfer_eq}, where the first step is performed by Faraday rotation depending only on $\mathbf{B}_\mathrm{los}$ and the fact that Stokes $V$ does not change with coordinate transformations in the plane of the sky. Changes of the orientation of $\mathbf{B}_\mathrm{sky}$ in time will only change the observer's definition of Stokes $Q$ and $U$. Indeed, this is likely to be the explanation for the observed wandering values of $Q$ and $U$ \citep{marr06b}, while Stokes $V$ appears to be stable.

\subsubsection{Relation to Observed Faraday Rotation Rates}\label{sec:depths}
A key assumption in the self-absorbed approximation above is that synchrotron emission, Faraday conversion, and Faraday rotation (which drives conversion)
all take place within the same narrow shell of plasma. Previous work \citep[e.g.,][]{marr07} has assumed that Faraday rotation primarily occurs far from the emission region at a given frequency. This ``Faraday screen'' approximation may still operate in regions of the accretion envelope where the material is too optically thin to contribute any
emission or absorption to the incoming radiation while still being dense and magnetized enough to produce Faraday rotation. For such cases, \citet{marr06a} calculate the cumulative Faraday rotation acting upon polarized radiation emitted at radius $r_0$ (analogous to eq.~\ref{eq:conversion}) as 
\begin{equation}\label{eq:rotation}
\Delta\chi(r_0,\nu)=\int_{r_0}^\infty \cfrac{2\pi}{c}\cfrac{\nu_p^2\nu_B}{\nu^2}\cos\theta~\Theta_\mathrm{FR}~dr~~,
\end{equation}
(except for the $\Theta_\mathrm{FR}$ term, which is taken to be 1). The total emission (Stokes $I$, $V$, and total LP) is assumed to come from a region interior to $r_0$ and to be essentially unaffected as it emerges from the source, except for the Faraday rotation of the electric vector position angle (EVPA). 

To be consistent with our self-absorbed Faraday conversion approach, Faraday rotation should act both internally (at $\tau\sim1$, where it drives conversion)
as well as externally (at $\tau\ll1$, where the Faraday screen approximation is valid). To satisfy this requirement, the
Faraday rotation scale lengths should extend far beyond the thickness of the emission/conversion shell. To confirm this, we
define the differential Faraday rotation and conversion depths (differential forms of eqs.~\ref{eq:conversion} and~\ref{eq:rotation}) as
\begin{equation}\label{eq:rotation_depth}
\frac{d\tau_\mathrm{FR}}{dr}\propto r^{-(\beta+\alpha)}\Theta_\mathrm{FR}(r)\nu^{-2}
\approx  r^{-(\beta+\alpha)} \frac{\ln{\gamma_\mathrm{rad}(r)}}{\gamma_\mathrm{rad}(r)}r_0^{2n}
\end{equation}
and
\begin{equation}\label{eq:conversion_depth}
\frac{d\tau_\mathrm{FC}}{dr}\propto r^{-(\beta+2\alpha)}\Theta_\mathrm{FC}(r)\nu^{-3},
\approx r^{-(\beta+2\alpha)} \gamma_\mathrm{rad}(r) r_0^{3n}~~,
\end{equation}
respectively, where $\gamma_\mathrm{rad}\propto r^{-\delta}$ (Section~\ref{sec:selfabsorbed}) is also a power law of $r$, with  $\delta$ small. It can be readily
seen that ${d\tau_\mathrm{FC}}/{dr}$ is significantly steeper in radius than ${d\tau_\mathrm{FR}}/{dr}$. For comparison, the synchrotron optical
depth can be written in differential form as ${d\tau}/{dr}\sim r^{-\beta-\alpha(p+2)/2}\nu^{-(p+4)/2}$ (see Section~\ref{sec:selfabsorbed}). For values of $p$ between 2 and 3, ${d\tau}/{dr}$ can be as steep as or steeper than ${d\tau_\mathrm{FC}}/{dr}$, and thus also steeper than ${d\tau_\mathrm{FR}}/{dr}$, confirming that significant Faraday rotation takes place outside of the $\tau=1$ surface \citep[see][]{jon77a}.  This means that Faraday rotation is still at work when Faraday conversion has stopped being effective. In other words,  the source has only a limited spatial range to produce Stokes $V$---driven by the sequence $Q\rightarrow U\rightarrow V$---before the conversion rate decreases to negligible values.  Therefore, while Faraday conversion is roughly local to the shell of radius $r_0$, Faraday rotation is not. 

A consequence of having significant Faraday rotation within and immediately outside the photosphere is a frequency-dependent rotation measure. The highest frequencies probe deeper layers in the stratified source and therefore integrate Faraday rotation contributions from the shells that define lower-frequency photospheres. There are, as of yet, few observational constraints on this possibility.
The RM has been determined through the measurement of a change in $\chi$ across a narrow range in frequency \citep{marr07}, and less securely through the comparison of position angles averaged over several observations at more widely separated frequencies \citep{mac06}.
A demonstration of the frequency dependence of the RM could be obtained through position angle measurements at multiple pairs of closely spaced frequencies, which should be possible very soon with improved millimeter interferometers and ALMA.

Although the Faraday screen approximation is reasonable for a stratified synchrotron source, it can be difficult
to obtain an accurate RM observationally, due to the source's layered geometry. Assuming the cold, optically thin Faraday rotation relation $\chi(\nu)=\chi_0+{c^2}/{\nu^2}\mathrm{RM}$ \citep[e.g.,][eq.~1]{marr06a}, RM is observationally obtained by computing the quantity $d\chi/d\lambda^2=d\chi/d(c/\nu^2)\approx \Delta \chi/ \Delta(c^2/\nu^2)$, using two neighboring frequencies and their respective observed EVPAs. The implied assumption, in addition to the validity of the screen, is that the emission at the two observed frequencies is originated in the same region and with the same initial EVPA $\chi_0$. However, in the layered scenario, if these two frequencies correspond to two shells that do not overlap spatially, they will produce emission that will travel through two different Faraday screens. In that case, RM would be a function
of the two frequencies observed and not a constant value. Thus far, observations suggest that this is not the case and that RM is indeed a constant, even when $\chi_0$ varies in time \citep{marr07}. Simultaneous observations at more than two frequencies will help resolve the question of whether Faraday rotation in \sgr\ is proportional to $\nu^{-2}$ or not.

One unaddressed problem in the discussion above is that there apparently is not enough observed LP at low frequencies, (e.g., 4 and 8 GHz, where CP dominates) to generate the observed Stokes $V$ via the Faraday conversion process. However, synchrotron flux density is highly polarized for ordered magnetic fields, even in the optically thick limit \citep{pac70}. Therefore,
it is highly likely that the observed LP suffers severe beam smoothing at these frequencies, at a point beyond the conversion region, if the Faraday-rotation region shows fluctuations (e.g., stochastic or turbulent in nature) at spatial scales smaller than the source size. For the sake of argument, consider partially polarized radiation with constant polarization angle $\chi_0$ entering a piecewise uniform Faraday screen. Radiation will exit the screen with rms fluctuations in EVPA equal to $\sigma_\chi\sim\sigma_\mathrm{RM}(c/\nu)^2$, where $\sigma_\mathrm{RM}$ contains the fluctuations in $n_e$, $B$, and $\theta$.  If the fluctuations in the outcoming $\Delta\chi$ are Gaussian, the observed LP will be $\langle m_l\rangle  = m_{l,0} \,e^{-2\sigma_\chi^2}$, where $m_{l,0}$ is the original LP fraction entering the screen. This occurs because, while $Q$ and $U$ are additive quantities, total LP is not, resulting in observed values of $m_l$ that are severely sensitive to EVPA changes in the plane of the sky.  

\subsection{The Faraday Conversion at Finite Temperatures}
\label{sec:FC_hot}
In an accretion flow, many of the physical quantities relevant for synchrotron emission are expected to increase inwardly. This is true for the electron density $n_e$ and the temperature $T$. In particular,
the temperature of the plasma near the event horizon can achieve mildly to highly relativistic temperatures. It is evident then that the cold plasma approximation should not apply in the interior regions where Faraday conversion could operate.  A more general expression for the Faraday transport coeffcients (eqs.~\ref{eq:kappaV} and \ref{eq:kappaQ}) is
\begin{eqnarray}\label{eq:cold_coef1}
\nonumber \kappa_V^*&=&\cfrac{2\pi}{c}\cfrac{\nu_p^2\nu_B}{\nu^2}\cos\theta
\, \Theta_\mathrm{FR}(\nu,T_e,\theta,B)\\
 &=&{\kappa_V^*}_\mathrm{cold}
\, \Theta_\mathrm{FR}(\nu,T_e,\theta,B)\\ \nonumber \\
\label{eq:cold_coef2}
\nonumber \kappa_Q^*&=&-\cfrac{\pi}{c}\cfrac{\nu_p^2\nu_B^2}{\nu^3}\sin^2\theta
\, \Theta_\mathrm{FC}(\nu,T_e,\theta,B)\\
 &=&{\kappa_Q^*}_\mathrm{cold}\, \Theta_\mathrm{FC}(\nu,T_e,\theta,B)~~,
\end{eqnarray}
where the functions $\Theta_\mathrm{FR}(\nu,T_e,\theta,B)$ and $\Theta_\mathrm{FC}(\nu,T_e,\theta,B)$
are finite-temperature corrections to the standard cold-plasma transport coefficients. By definition,
$\Theta_\mathrm{FR,FC}\rightarrow1$ when $T\rightarrow0$ and  $\nu\gg\nu_B$.  

In a cold plasma, the electrons oscillate in response to electromagnetic waves but are effectively stationary otherwise. When the electron energy distribution is nonnegligible, the interaction of electron motions and incident electromagnetic radiation is given by the Boltzmann equation in the presence of the Lorentz force: the Vlasov equation. The Vlasov equation describes the dielectric tensor of the plasma and thus its Faraday transport coefficients, which depend on the electron energy distribution \citep[e.g.,][]{mel05}.

One of the earliest approaches to this problem originated with \citet{saz69}. For a power-law distribution of electron energies, the ratio of the Faraday transport coefficients \citep[see also][]{jon77a} is
\begin{equation}\label{eq:pl_ratio}
\begin{split}
\left|\frac{\kappa_Q^*}{\kappa_V^*}\right|_\mathrm{pl}
= &\left|\frac{\kappa_Q^*}{\kappa_V^*}\right|_\mathrm{cold}\frac{\Theta_\mathrm{FC}^\mathrm{pl}}
{\Theta_\mathrm{FR}^\mathrm{pl}}=\left|\frac{\kappa_Q^*}{\kappa_V^*}\right|_\mathrm{cold}
 \\ &\times 
 \frac{2(p+1)}{(p+2){(p-2)}}
\cfrac{\gamma_\mathrm{min}^{3}}{\ln \gamma_\mathrm{min}}
\left[1-\left(\gamma_\mathrm{min}^2\cfrac{\nu_{B\perp}}{\nu}\right)^{\frac{(p-2)}{2}}\right]~~,
\end{split}
\end{equation}
which is an increasing function of $\gamma_\mathrm{min}$ (\hbox{$\gamma_\mathrm{min}>1$} always) under the condition \hbox{$\nu_B/\nu\ll1$}. Eq.~\ref{eq:pl_ratio} shows that even with $\left(\kappa_Q^*/\kappa_V^*\right)_\mathrm{cold}\ll1$, for
hot enough plasmas, the Faraday conversion coefficient $\kappa_Q^*$ can become a significant fraction of $\kappa_V^*$ or even exceed it in extreme cases.
 
For electrons in a relativistic thermal (Maxwell-Juttner) distribution, the plasma dielectric tensor is different. 
\citet{mel97} derived an ultrarelativistic approximation for the response tensor of a magnetized thermal plasma. 
 \citet{bal07} combined this result with that of a classical cold plasma to obtain an approximate continuous function of the polarization proper modes as a function of temperature.
\citet{shch08} extended these results, providing consistent analytic expressions valid for all temperatures. In this case, the ratio of the Faraday transport coefficients is
\begin{equation}\label{eq:thermal_ratio}
\begin{split}
\left|\frac{\kappa_Q^*}{\kappa_V^*}\right|_\mathrm{therm}=&\left|\frac{\kappa_Q^*}{\kappa_V^*}\right|_\mathrm{cold}\frac{\Theta_\mathrm{FC}^\mathrm{therm}}
{\Theta_\mathrm{FR}^\mathrm{therm}}=\left|\frac{\kappa_Q^*}{\kappa_V^*}\right|_\mathrm{cold}
\left[\cfrac{K_1\left(\frac{m_ec^2}{k_BT}\right)}{K_0\left(\frac{m_ec^2}{k_BT}\right)}\right. \\ & \left.+
6\frac{k_BT}{m_ec^2}\cfrac{K_2\left(\frac{m_ec^2}{k_BT}\right)}{K_0\left(\frac{m_ec^2}{k_BT}\right)}\right]
\frac{g\left(\frac{}{}T,\theta,\nu_B/\nu\right)}{f\left(\frac{}{}T,\theta,\nu_B/\nu\right)}~~,
\end{split}
\end{equation}
where $K_n(x)$ is the modified Bessel function of the second kind.  Eq.~\ref{eq:thermal_ratio} smoothly reduces to the cold plasma limit when $T\rightarrow0$, since the term in the square brackets becomes exactly 1 in this limit. At temperatures of $10^9$K ($kT/m_ec^2$ = 1/6), the
term in the square brackets takes a value of $\sim2.5$, and at $2\times10^{11}$K ($kT/m_ec^2$ = 34), it reaches $\sim1.3\times10^5$.

The factors $f$ and $g$, above, introduced by \citet{shch08}, take a value of 1 for $\nu_B/\nu\rightarrow0$ and provide the necessary corrections for higher-order terms in $\nu_B/\nu$. They can lead to significant modifications to the ratio of the Faraday transport coefficients in the moderately relativisitic regime.
For example, for intense magnetic fields, the cyclotron frequency can be written $\nu_B=0.35\,\mathrm{GHz}\left(B/20\mathrm{G}\right)$, and thus the ratio $\nu_B/\nu$ takes values of $\sim10^{-3}$ at submillimeter wavelengths. Following \citet{shch08}, at  $\nu_B/\nu=10^{-3}$ and $T=2\times10^{11}$~K with $\theta=\pi/4$, the correction factors become $g\sim0.93$ and $f\sim-0.003$. The multipliers of the cold Faraday transport coefficients in eq.~\ref{eq:thermal_ratio} can strongly change the expected behavior even at moderate temperatures. This clearly demonstrates the importance of including these factors in numerical simulations of \sgr\ in order to properly handle the PRT problem.

\section{Conclusions}\label{sec:conclusion}

We have reported the first detection of circularly polarized radiation toward Sagittarius A* at submillimeter wavelengths. The detected CP fractions are at levels of $1.2\pm0.3\%$ and  $1.6\pm0.3\%$ for 230 GHz and 345 GHz, respectively, and are confirmed by observations at different epochs. We have tested the significance of this detection by analyzing the gain calibration, spectral variability and time variability, and by observing other sources. CP is not found in any other source, with the most sensitive limits confining uncalibrated instrumental CP to less than 0.2\%. 

Our measurements of CP, combined with previously reported measurements at lower frequencies, indicate that the polarization fraction rises with frequency. The sign (handedness) of the CP signal is the same for all detections at all frequencies over the period 1981 to 2007, as was found in low-frequency measurements alone \citep[e.g.,][]{bow02}. The average CP fraction as a function of frequency is remarkably well characterized by a power law with $\nu^{0.35\pm0.03}$. However, there have been no detections of CP between 15 and 230 GHz, so the CP spectrum may not be monotonic.

The general trend of slowly increasing CP with frequency is consistent with self-absorbed Faraday conversion. We discuss a simple set of power-law scalings of $B$ and $n_e$ that allow Faraday conversion to operate with only weak frequency dependence, imposing no CP sign reversals as long as the magnetic field itself does not reverse over the range of radii providing centimeter to submillimeter emission. In contrast, more sophisticated models in the literature \citep{hua08,hua09} show frequent reversals in submillimeter CP and therefore are not consistent with the results presented here. 

If the CP originates from self-absorbed Faraday conversion, the corresponding Faraday rotation depth could vary with frequency, which would be contrary to the assumption of a cold Faraday screen \citep{mac06,marr07}. Polarization observations in the near future will test the frequency variation of the RM, and there already exists a claim of an RM change between 22 and 230 GHz \citep{yus07}. 
In general, the PRT in a self-absorbed synchrotron source should couple the variations in LP and CP, so sensitive observations of variability in the polarization of this source will be crucial to understanding the structure of the emission region and the source of CP. As with variations in the RM \citep{sha07,pan10}, variations in CP will occur on a timescale set by the structure of the accretion flow and can therefore differentiate between models.

{\bf Acknowledgments.}
We thank Avery Broderick, Siming Liu, and Roman Shcherbakov for helpful comments and discussions. Support for DPM was provided by NASA through Hubble Fellowship grant HST-HF-51259.01 awarded by the Space Telescope Science Institute, which is operated by the Association of Universities for Research in Astronomy Inc. for NASA, under contract NAS 5-26555.

{\it Facilities:} \facility{SMA (Polarimeter)}


\input{circular_sgrastar_revision.bbl}
 \clearpage

\end{document}

%% file: table_pol_b.tex
\begin{deluxetable*}{cccc}
\pdfoutput=1
\tabletypesize{\scriptsize}
\tablecolumns{6} 
\tablewidth{0pc} 
\tablecaption{Radio-to-Submillimeter Linear and Circular Polarization of Sgr A*
\label{tab:table_pol}}  
\tablehead{ 
\colhead{Frequency (GHz)}  & 
\colhead{Fractional LP ($\%$)\tablenotemark{a}}  &\colhead{Fractional Stokes $V$ ($\%$)\tablenotemark{b}}  &\colhead{Reference}}
\startdata 
1.4 &\nodata & $-0.21\pm0.10$ (13)&\citet{bow02}\\
4.8 &\nodata&$-0.33\pm0.07$ (13)&''\\
8.4 &\nodata&$-0.32\pm0.08$ (13)&''\\
15 &\nodata&$-0.62\pm0.26$ (12)&''\\
 \cline{1-4}
4.8 &  \nodata  & $-0.37\pm0.04$ & \citet{sau99} \\ 
 \cline{1-4}
4.8 & $<0.08$ & \nodata &\citet{bow99a}\\
8.4 & $<0.17$ & \nodata & ''\\
 \cline{1-4}
4.8&\nodata & $-0.36\pm0.05$ (3)&\citet{bow99b}\\
8.4 &\nodata &$-0.26\pm0.06$ (3)&''\\
 \cline{1-4}
22 & $<0.2$ &\nodata &\citet{bow99c}\\
43 & $<0.4$&\nodata&''\\
86 & $<1.0$&\nodata&''\\
 \cline{1-4} 
 82.8& $2.1\pm0.4$ &\nodata &\citet{mac06}\\
86.3 &$0.8\pm0.5$&\nodata&'' \\
 \cline{1-4} 
 100&\nodata&$<2.0$ &\citet{tsu03}\\
 \cline{1-4} 
112& $<3.6$&$<3.6$ &\citet{bow01}\\
 \cline{1-4}
216&$9.1\pm2.2$&\nodata &\citet{bow05}\\
230 &$10.0\pm2.5$&\nodata&''\\
 \cline{1-4} 
 230&$7.2\pm0.6$ &$\lesssim2$ &\citet{bow03}\\
 \cline{1-4} 
 230& $5.9\pm1.6$& \nodata&\citet{marr07}\\
 \cline{1-4} 
340& $6.4\pm2.0$&$\lesssim1$ &\citet{marr06a}\\
 \cline{1-4} 
 150 & $12_{-4}^{+9}$ & \nodata&\citet{ait00}\\
225& $11_{-2}^{+3}$&\nodata&''\\
350 &$13_{-4}^{+10}$&\nodata&''\\
400 &$22_{-9}^{+25}$&\nodata&''\\
\enddata 
\tablenotetext{a}{Uncertainties can correspond to both systematic errors and time-variability dispersion. For example, data corresponding to the multi-epoch observations of \citet{marr06a,marr07} are presented with errors corresponding to the standard deviation of the sample.}

\tablenotetext{b}{The errors shown are the standard deviation (rms) of the sample, not the standard deviation of the mean. The number of samples is listed in parentheses. For the \citet{bow02} data,
we show the results for the 1999 VLA measurements only because of the advanced calibration technique used. However, they are consistent with the archival VLA and the ATCA data also reported by \citet{bow02}. Upper limits are shown at the 2-$\sigma$ level.}

\end{deluxetable*}

%% file: tab_tracks2.tex
\begin{deluxetable}{lrrcr} 
\tabletypesize{\scriptsize}
\tablecolumns{5} 
\tablewidth{0pc} 
\tablecaption{Observation Epochs\label{tab:tracks2}} 
\tablehead{ 
 \colhead{Date}  & \colhead{Main}    & \colhead{Freq.\tablenotemark{a}}  &
\colhead{SMA}     & \colhead{$\tau_\mathrm{225\,GHz}$}   \\
 \colhead{}  & \colhead{Target}    & \colhead{[GHz]}  &
\colhead{Config.\tablenotemark{b} }     & \colhead{}   
}
\startdata 
2005 June 6&Sgr~A*&343.0&CN&0.055 \\
2006 July 17&Sgr~A*&226.9&VEX&0.05-0.08 \\
2007 March 31&Sgr~A*&226.9&CN& 0.055 \\
2008 May 30&1924-292&226.9&CN&0.08 
\enddata 
\tablenotetext{a}{Frequency of the local oscillator. Upper and lower sidebands are centered 5 GHz above and below this frequency, respectively.}
\tablenotetext{b}{Array configurations include ``Compact North" (CN) and ``Very Extended" (VEX).}
\end{deluxetable}

%% file: tab_sgra_pol.tex
\begin{deluxetable*}{lcrrrr}
\pdfoutput=1
\tablecaption{Double-Sideband Flux and (Fractional) Polarization for Sgr A*\label{tab:sgra_pol}} 
\tablehead{ \colhead{Date} &  \colhead{Freq. [GHz]} & 
 \colhead{$I$\tablenotemark{a} [Jy]}    & \colhead{$Q/I$} & 
\colhead{$U/I$}    & \colhead{$V/I$}  }
\startdata 
2005 June 6&343.0&$3.17\pm0.02$&$0.027\pm0.003$&$-0.049\pm0.003$&$-0.016\pm0.003$\\
2006 July 17&226.9& $3.88\pm0.02$ &$0.020\pm0.003$ &$ -0.053\pm0.003$&$-    0.011\pm0.003$ \\
2007 March 31&226.9&$3.52\pm0.01$ &$-0.061\pm0.001$&$-0.032\pm0.001$&$-0.012\pm0.001$ \\
2008 May 30\tablenotemark{b}&226.9&$3.87\pm0.02$&$0.044\pm0.002$&$-0.032\pm0.002$&$-0.011\pm0.004$ \\
\enddata 
\tablenotetext{a}{Statistical error only. Absolute calibration precision is typically 10-20\%.}
\tablenotetext{b}{Sgr~A* was observed for one hour prior to the 1924-292 test observation.}
\end{deluxetable*}

%% file: table_offsets.tex
\begin{deluxetable*}{cccc} 
\pdfoutput=1
\tablewidth{305pt} 
\tablecaption{Circular Polarization in Subsections of March 31, 2008 Track\label{tab:offsets}} 
\tablehead{ \colhead{UT hour range}  & \colhead{$V/I$}   & \colhead{$\Delta\theta_V$} & \colhead{SNR$_V$}  \\
\colhead{}  & \colhead{}   & \colhead{(arcsec)}    & \colhead{}}
\startdata 
 12.4 -14.2 &$(-1.3\pm0.2)\times10^{-2}$  &  1.52  $\pm$ 0.27    &   6.5\\
14.2 -15.7 & $(-1.1\pm0.1)\times10^{-2}$    &  0.78  $\pm$ 0.21    &   7.8\\
 15.7 -17.2 &  $(-1.6\pm0.1)\times10^{-2}$ &  0.77  $\pm$ 0.10    &  10.5\\
17.2 -18.9 &  $(-1.6\pm0.2)\times10^{-2}$ &  0.31  $\pm$ 0.18    &   9.1
\enddata 
\end{deluxetable*} 

%% file: tab_quasars_pol.tex
\begin{deluxetable*}{lrrcrr} 
\pdfoutput=1
\tabletypesize{\scriptsize}
\tablecolumns{6} 
\tablewidth{0pc} 
\tablecaption{Circular Polarization\tablenotemark{a} for Test Quasars on March 31, 2007\label{tab:quasars_pol}} 
\tablehead{ 
&\multicolumn{2}{c}{LSB}&&\multicolumn{2}{c}{USB}\\
\cline{2-3}\cline{5-6}\\
 \colhead{Source}   &\colhead{$I$ [Jy]} & \colhead{$V/I$} & &
\colhead{$I$ [Jy]} & \colhead{$V/I$}
}
\startdata 
3C273\tablenotemark{b}&
$15.40\pm0.03$&$(1.2\pm1.2)\times10^{-3}$&&$15.05\pm0.02$&$(1.1\pm1.3)\times10^{-3}$\\
3C279&$12.92\pm0.02$&$(1.5\pm1.4)\times10^{-3}$&&$12.88\pm0.02$&$(1.3\pm1.4)\times10^{-3}$\\
3C286&$0.49\pm0.01$&$(7.9\pm15.2)\times10^{-3}$&&$0.46\pm0.01$&$(14.8\pm17.9)\times10^{-3}$\\
1337-129&$6.92\pm0.03$&$(2.1\pm3.0)\times10^{-3}$&&$6.89\pm0.03$&$(2.5\pm3.4)\times10^{-3}$ \\
\hline
1733-130\tablenotemark{c}&
$1.48\pm0.04$&$ (0.8\pm0.4)\times10^{-3}$&&$1.46\pm0.04$&$(1.0\pm0.5)\times10^{-3}$\\
\hline
1924-292\tablenotemark{d}&$7.00\pm0.01$&$(-0.2\pm1.1)\times10^{-3}$&&
$6.95\pm0.01$&$(-0.1\pm1.2)\times10^{-3}$
\enddata 
\tablenotetext{a}{Stokes $V$ fluxes were calculated by fitting the visibility data to visibilities corresponding to a point source located at the interferometric (phase) center of the map.}
\tablenotetext{b}{3C273 was calibrated in polarization and gain with 3C279. In contrast, quasars 3C279, 3C286, and 1337-129 were calibrated using 3C273 as a gain and a polarization calibrator.}
\tablenotetext{c}{Quasar 1773-130 is the gain calibrator for Sgr A*. The average gain curve was derived using both $LL$ and $RR$ visibilities, then applied to the 1733-130 data before measuring the CP.}
\tablenotetext{d}{Shown here for comparison, quasar 1924-292 was observed on the night of May 30, 2008.}
\end{deluxetable*} 